\documentclass[manuscript]{aastex}

\slugcomment{Submitted to ApJ}

\shorttitle{Heliospheric Evolution of Magnetic Clouds}
\shortauthors{Vr\v{s}nak et al.}

\begin{document}

\title{Heliospheric Evolution of Magnetic Clouds}

\author{B. Vr\v{s}nak\altaffilmark{1} }
\email{bvrsnak@geof.hr}

\author{T. Amerstorfer\altaffilmark{2}, M. Dumbovi\'{c}\altaffilmark{3}, M. Leitner\altaffilmark{3}, A.M. Veronig\altaffilmark{3,4}, M. Temmer\altaffilmark{3}, C. M\"ostl\altaffilmark{2},
U.V. Amerstorfer\altaffilmark{2}, C.J. Farrugia\altaffilmark{5}, A.B. Galvin\altaffilmark{5}}



\altaffiltext{1}{Hvar Observatory, Faculty of Geodesy, University of Zagreb, Ka\v{c}i\'{c}eva 26, HR-10000 Zagreb, Croatia}

\altaffiltext{2}{Space Research Institute, Austrian Academy of Sciences, Graz, Austria}

\altaffiltext{3}{Institute of Physics, University of Graz, Austria}

\altaffiltext{4}{Kanzelh\"ohe Observatory for Solar and Environmental Research, University of Graz, Austria}

\altaffiltext{5}{Space Science Center and Department of Physics, University of New Hampshire, Durham, NH 03824, USA}

\begin{abstract}
Interplanetary evolution of eleven magnetic clouds (MCs) recorded by at least two radially aligned spacecraft is studied. The \emph{in situ} magnetic field measurements are fitted to a cylindrically symmetric Gold-Hoyle force-free uniform-twist flux-rope configuration. The analysis reveals that in a statistical sense the expansion of studied MCs is compatible with self-similar behavior. However, individual events expose a large scatter of expansion rates, ranging from very weak to very strong expansion. Individually, only four events show an expansion rate compatible with the isotropic self-similar expansion. The results indicate that the expansion has to be much stronger when MCs are still close to the Sun than in the studied 0.47\,--\,4.8\,AU distance range.
The evolution of the magnetic field strength shows a large deviation from the behavior expected for the case of an isotropic self-similar expansion. In the statistical sense, as well as in most of the individual events, the inferred magnetic field decreases much slower than expected. Only three events show a behavior compatible with a self-similar expansion. There is also a discrepancy between the magnetic field decrease and the increase of the MC size, indicating that magnetic reconnection and geometrical deformations play a significant role in the MC evolution.
About half of the events show a decay of the electric current as expected for the self-similar expansion.
Statistically, the inferred axial magnetic flux is broadly consistent with it remaining constant. However, events characterized by large magnetic flux show a clear tendency of decreasing flux.
\end{abstract}


\section{Introduction}

Eruptions of unstable magnetic structures in the solar atmosphere result in the occurrence of the so-called coronal mass ejections (CMEs), usually observed by white light coronagraphs. Their interplanetary counterparts (ICMEs) frequently show structures that have characteristics of a helically twisted flux rope
\citep[for a historical background see, e.g.,][]{burlaga81,klein82,burlaga88,gosling90,lepping90,B&S98}, usually denoted as magnetic clouds \citep[for terminology see, e.g.,][and references therein]{burlaga88,rouillard11,mostl12apj}. Throughout the paper the term magnetic cloud (MC) will be used exclusively for the flux-rope element of the ICME, whereas the term ICME will be used for the overall structure of an ejection, including the shock and sheath region.

Early stages of the eruption are governed by the Lorentz force that accelerates the CME and causes its rapid expansion \citep[e.g.,][and references therein]{vrs08angeo,chen10}. After the main acceleration phase, at larger heliocentric distances the Lorentz force ceases \citep{vrs04domi} and the evolution of ICMEs becomes dominated by the interaction of the ICME with the ambient solar wind, resulting in several significant effects. First, the overall dynamics is affected by ``MHD/aerodynamic'' drag \citep[e.g.,][and references therein]{cargill04,vrs08dij,vrs13dbm}, causing deceleration/acceleration of ICMEs that are faster/slower than the ambient solar wind, i.e., the kinematics of the ICME and the embedded MC tend to adjust to the solar wind flow \citep[e.g.,][]{gopal00}. Second, the MC expansion in the radial direction weakens with heliocentric distance \citep[e.g.,][]{leitner07,gulisano12}, leading to a deformation of the MC cross section. As a matter of fact, numerical simulations show that MCs should attain a convex-outward shape due to the pressure gradients \citep[``pancaking effect''; see, e.g.,][]{cargill94,cargill96,cargill99,hidalgo03,riley04,farrugia05,liu06,owens06,ruffenach15}.
   In this respect, it should be noted that such a scenario is basically coming from a two-dimensional (2D) approach, and it could be significantly altered in more realistic 3D simulations. 

Under suitable conditions, there is another effect that might play a significant role in the CME evolution. Namely, magnetic reconnection of the MC magnetic field and the ambient interplanetary magnetic field might occur, reducing the MC magnetic flux and the MC cross-section area by ``peeling-off'' the outer layers of the flux rope \citep[for the latter see, e.g.,][]{dasso06,dasso07,gosling07,mostl08,ruffenach12}, as well as causing a deflection of the MC motion \citep[][]{cargill96,vandas96,ruffenach15}.
However, this effect was mostly inferred from rather simple models, such as the Lundquist constant-alpha force-free 1D configuration. Therefore,  the interpretations based on such  a simplified approach, and especially the quantitative estimates, should be taken with caution.
Finally, let us mention that if taking place within the MC, reconnection can significantly change its internal structure \citep[e.g.,][]{farrugia01,gosling05,gosling07}.

A significant point is that the \emph{in situ} measurements clearly show that most of the MCs expand relative to the ambient solar wind, since the plasma speed at the MC front is significantly higher than at its rear \citep[e.g.,][]{klein82,farrugia93,lepping03,dasso07,demoulin08,lepping08,demoulin09,rouillard09,gulisano10,gulisano12}.
The evolutionary aspect of the MC expansion was investigated with various approaches: (i) employing multispacecraft \emph{in situ} measurements in a radially aligned configuration \citep[e.g.,][]{burlaga81,burlaga82sph,osherovich93,B&S98,mulligan01grl}, (ii) inspecting remote observations \citep[e.g.,][]{rouillard09,savani09}, or (iii)
applying a statistical approach, i.e., investigating sizes of a number of MCs as a function of heliocentric distance \citep[e.g.,][]{kumar96,B&S98,liu05,wangC05,leitner07,gulisano10,gulisano12}.
  However, although the majority of MCs expand, it is important to note that \citet{jian18} have shown that at 1\,AU about 23\,\% of ICMEs do not expand, and that about 6\,\% of ICMEs (mostly slow ones)  even show contraction. The MC compression  in radial direction was also reported by  \citet{hu17}.
Note that the statistical approach was used also to infer the evolution of some other physical parameters of MCs, e.g., density, temperature and magnetic field strength \citep[e.g.,][]{kumar96,B&S98,liu05,farrugia05,wangC05,gulisano10,gulisano12,winslow15}.

In order to better understand the internal structure of MCs, their magnetic field configuration was modeled by a number of authors, either from a purely theoretical point of view, or by fitting various presumed magnetic field configurations to the \emph{in situ} measurements
\citep[e.g.,][]{burlaga88,lepping90,osherovich93,kumar96,B&S98,chinchilla02,cid02,hidalgo02,hu02,hidalgo03,romashet03,vandas03,chinchilla05,dasso06,dasso07,marubashi07,mostl09sph,mostl12apj,hu14}.
In most of studies where the magnetic structure of MCs was modeled by fitting to the \emph{in situ} measurements,
the data were gathered by a single spacecraft located at a given heliocentric distance, thus not providing information on the evolution of the magnetic structure of the analyzed MCs along their trajectory.

The most direct insight into the evolution of the internal magnetic structure of MCs can be gained by analyzing \emph{in situ} measurements of radially aligned and widely separated spacecraft (hereafter, ``aligned events''). Unfortunately, not too many aligned events are reported \citep[see, e.g., the lists provided by][]{leitner07,winslow15}, and only a handful of them were analyzed in detail. In this respect, it should be noted that some of the studies concerning aligned events were focused mainly on the analysis of the overall MC dynamics, whereas the internal structure was described only in the most basic terms \citep[e.g.,][]{mostl11,rollett14,amerstorfer18}.
In some other papers, where the MC evolution was studied statistically using samples of MCs observed over a wide range of heliocentric distances, the aligned events were briefly described for purposes of illustration, concentrating mainly on, e.g., the MC expansion, shock/sheath evolution, magnetic field strength, or overall dynamics \citep[e.g.,][]{burlaga81,B&S98,farrugia05,forsyth06}. In some of the aligned events studies, the separation of spacecraft that recorded the MC was not large enough to provide reliable information on the evolution of its internal structure \citep[e.g.,][]{burlaga82sph,rouillard09}.

To the best of our knowledge, in the last 25 years only ten papers fully devoted to the in-depth analysis of the evolution of the internal MC structure using data from sufficiently-separated spacecraft were published.
In the first papers of this kind, the data gathered by spacecraft Helios 2, Advanced Composition Explorer (ACE), The Near Earth Asteroid Rendezvous (NEAR), Ulysses, and Voyager 2, were used to infer the MC evolution beyond the 1\,AU \citep[][]{osherovich93,mulligan01grl,du07,nakwacki11}. Later on, after the  STEREO-A/STEREO-B  (hereafter, STA/STB), MESSENGER  (hereafter, MES), and Venus Express (hereafter, VEX) were launched, new data from these spacecraft were employed to get a radially-aligned measurements that provided information on the MC evolution within the Sun-Earth space  \citep[][]{chinchilla12,chinchilla13,good15,good18,kubicka16,wangY18}.

The aim of this paper is to contribute to the comprehension of heliospheric evolution of the internal structure of MCs by adding a detailed analysis of eleven aligned events recorded over the heliocentric distances from $\sim$\,0.5 to $\sim$\,5 AU. The study is focused on the evolution of the MC size and the magnetic field strength, which allows us also to infer the evolution of the axial magnetic flux and electric current.
The results are compared with previous studies, and in addition, the main shortcomings of the applied approach are identified.


\section{Measurements and Data Analysis}

\begin{figure}
\plottwo{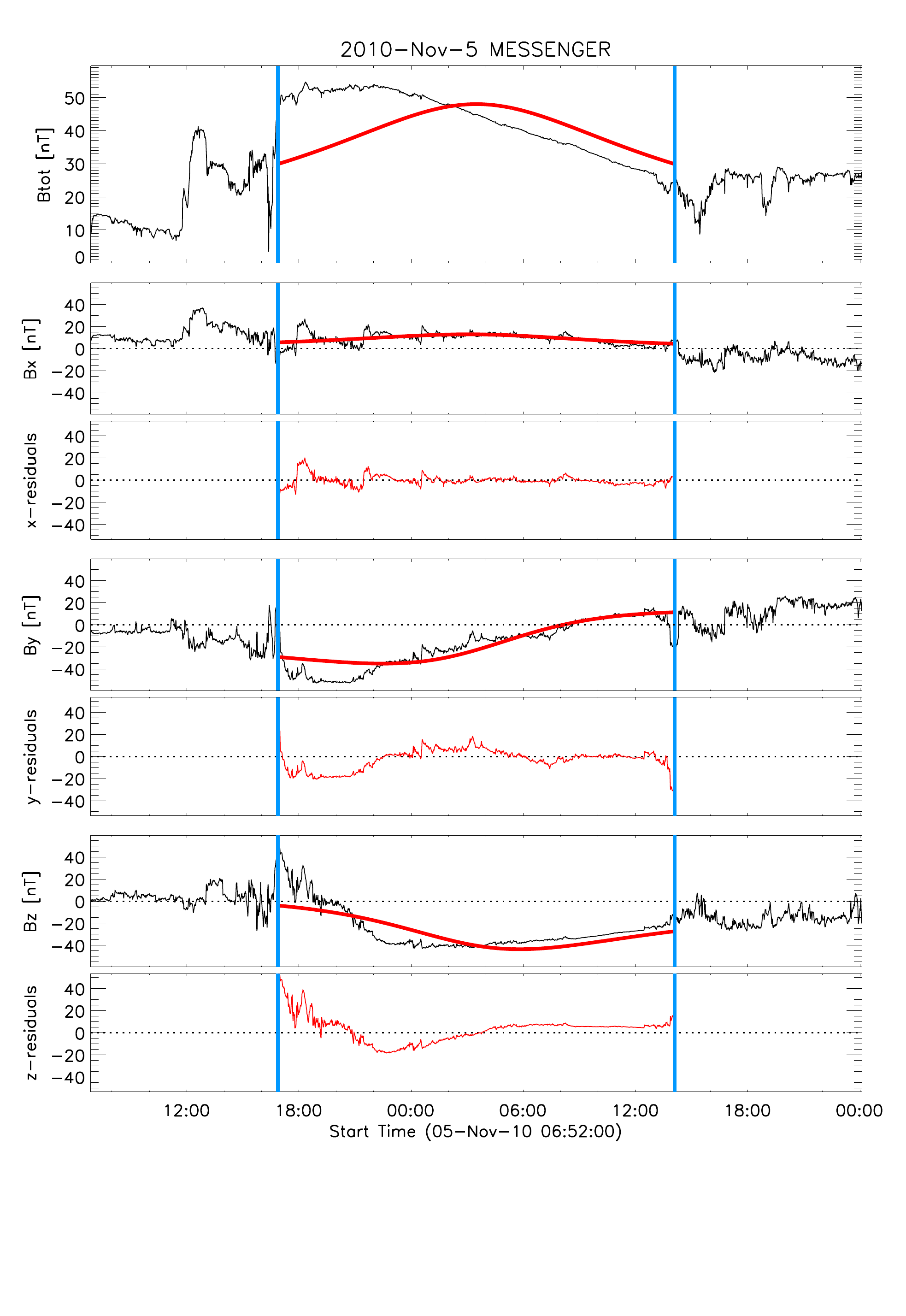}{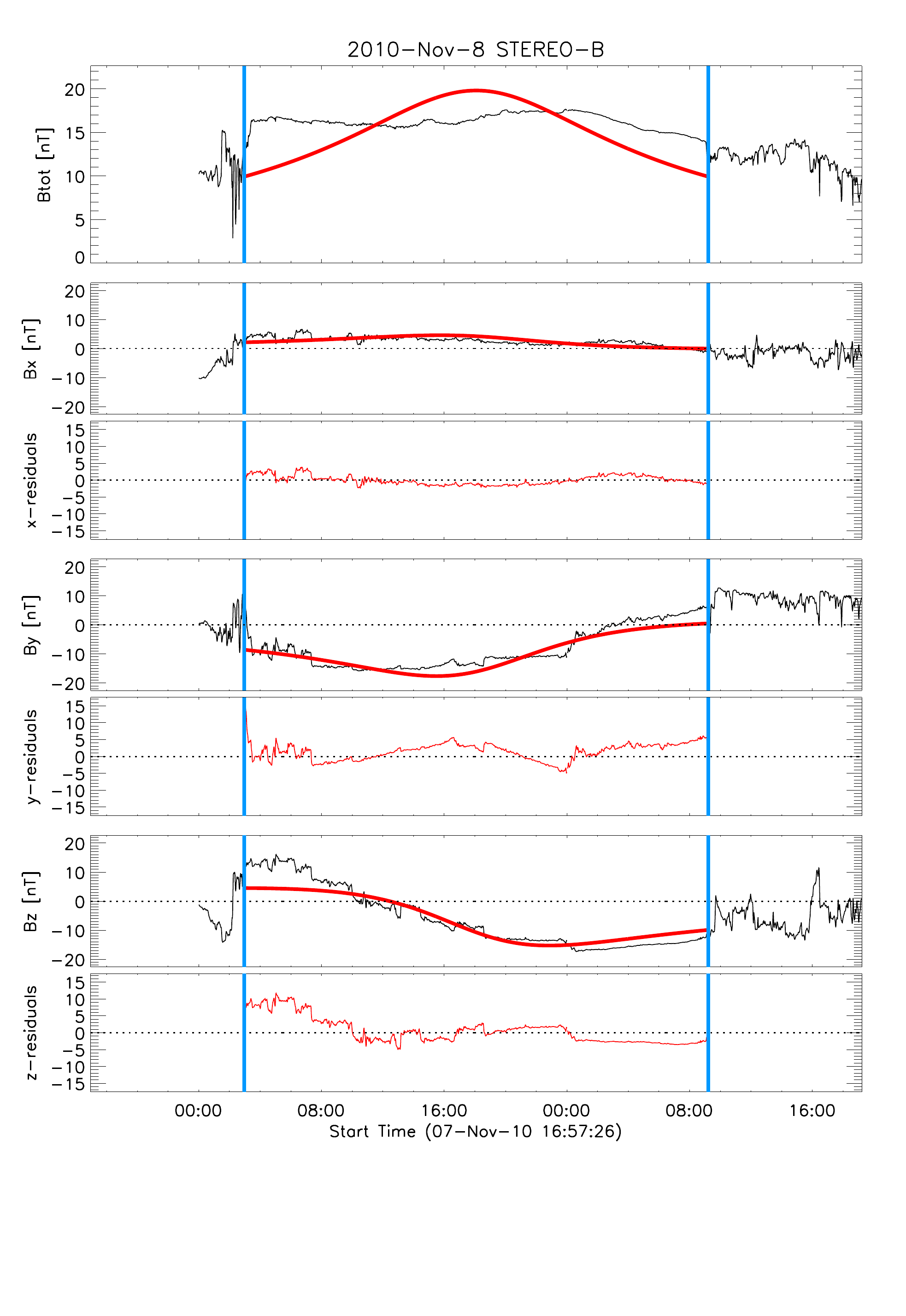}
  \caption
  {Example of the Gold-Hoyle flux-rope fitting (red curve): the event E8 recorded by MES on 2010 November 5 (left), and by STB on 2010 November 7/8 (right). The magnetic field strength $B_{tot}$ is given in the top panel, whereas the next six panels display magnetic field components and the residuals, respectively.
  \label{f00}
  }
\end{figure}

\begin{figure}
\plotone{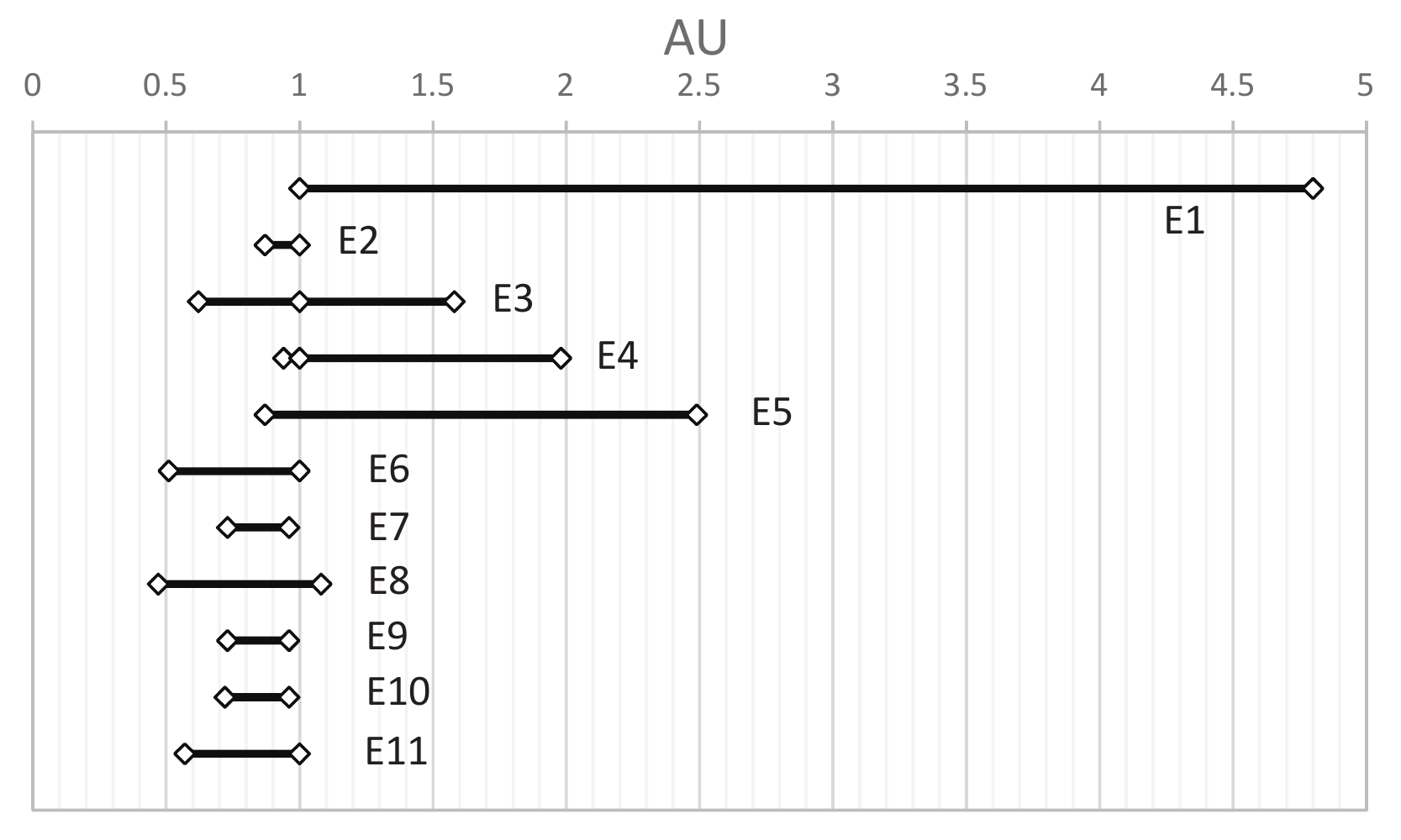}
  \caption
  {Distance ranges covered by the \emph{in situ} measurements. Diamond symbols mark positions of spacecraft.
  \label{f0}
  }
\end{figure}

In the following, we analyze 11 events that were observed by at least two radially aligned spacecraft.
In particular, we employ data measured by MESSENGER (MES), Venus Express (VEX), Helios 1 and 2 (H1, H2), Interplanetary Monitoring Platform-8 (IMP8), Wind, STEREO-A (STA), STEREO-B (STB), Pioneer 11 (P11), Voyager 1 and 2 (V1, V2), and compiled in The Space Physics Data Facility OMNI2 data base providing spacecraft-interspersed, near-Earth solar wind data ({\tt http://omniweb.gsfc.nasa.gov/}). An example of the measurements by two radially aligned spacecraft is presented in Figure~\ref{f00}. The list of events is displayed in Table~\ref{tab1}, where Column 1 gives the data label, Columns 2 and 3 the year and the data sources, and Column 4 the time range (expressed in Day of Year; DOY). In Column 5 the distance range (expressed in AU) covered by the measurements is presented; the measurements stretch from $R=0.47$ to 4.8 AU. The shortest distance between two spacecraft was in Event 2 (E2; 0.87\,--\,1.00 AU). It is included in this paper only to illustrate that if two spacecraft are too close, the results regarding the evolutionary behavior can become unreliable. It should be noted that the events E7, E9, and E10 were also measured by relatively closely positioned spacecraft ($\Delta R<0.35$ AU; see Table~\ref{tab1} and Figure~\ref{f0}). The largest distance range was in Event 1 (1.00\,--\,4.80 AU). Events E3 and E4 were measured by three spacecraft, where in E4 two of the three spacecraft were quite close ($R=0.94$ and 1), which we use to get an independent measure of the uncertainty of data obtained at a given distance. The distance ranges covered by measurements are presented for all events in Figure~\ref{f0}.
   Events 1\,--\,5 have already been used in the statistical study by \citet{leitner07}; the event labels from that study are given in Column 6.
Events 2 and 4 were also analyzed by \cite{farrugia05} (the event labels from that study are given in Column 6 in brackets) and
   Event 8 was analyzed by \citet{good18} and \citet{amerstorfer18}.

In all events, the magnetic field vector \textbf{\it B} was measured by all spacecraft. On the other hand, the plasma measurements, including the flow speed, are not available for the events measured by MES and VEX (Events 6\,--\,11). For these events we estimated the MC propagation speed indirectly, utilizing the following observational information:
\begin{itemize}
  \item the time $t_1$ when the CME was at the heliocentric distance of $R_1=20$ solar radii ($R_1=20\,R_S$) and its speed $v_1$ at this location \citep[estimated from coronagraphic measurements provided in the SOHO/LASCO CME catalog;][]{yashiro04};
  \item the time $t_2$ when the MC arrived to MES or VEX (the heliocentric distance $R_2$);
  \item the  arrival time $t_3$ and speed $v_3$ of the MC at the spacecraft located at $R_3\sim$1\,AU  (Wind, STA, or STB).
\end{itemize}

The transit time $\Delta t_{12}=t_2-t_1$ from $R_1$ to $R_2$  was used  to estimate the corresponding mean speed $v_{12}= (R_2-R_1)/\Delta t_{12}$, which was attributed to the half-distance $R_{12}=(R_1+R_2)/2$. In the same manner, we estimated the speed $v_{23}$ at the half-distance $R_{23}=(R_2+R_3)/2$. In this way, the MC speed at four heliocentric distances ($R_1$, $R_{12}$,  $R_{23}$, and $R_3$) was obtained. Finally, the four speed--distance data points $v_1(R_1)$, $v_{12}(R_{12})$, $v_{23}(R_{23})$, and $v_3(R_3)$, were used to interpolate the value of the MC speed at the location $R_2$ where MES or VEX were located.

The flux-rope magnetic field structure was reconstructed by fitting the \emph{in situ} magnetic field measurements with the Gold-Hoyle force-free uniform-twist configuration
\citep[][for details see Appendix A; for the fitting procedure see \citeauthor{farrugia99} \citeyear{farrugia99} and  \citeauthor{wangJ16} \citeyear{wangJ16};  
  for different reconstruction techniques see \citeauthor{alhaddad18} \citeyear{alhaddad18}; for the observational aspect see \citeauthor{vrs88} \citeyear{vrs88}]{gold60}.
 This model was chosen since it does not restrict the pitch angle of the field lines at the flux-rope boundary to $90^\circ$ as is the case of the frequently used Lundquist model \citep[][]{lundquist50}.
The fitting provides the magnetic field strength in the MC center, $B_c$, the latitudinal ($\theta$) and longitudinal ($\phi$) direction of the flux-rope axis ($\theta$ is ranging from $-90^{\circ}$ to 90$^{\circ}$, and $\phi$ is defined counterclockwise from positive x-direction, pointing towards the Sun and ranging between 0 and $360^{\circ}$), the impact parameter, $p$, and the sign of helicity, $H$.
The ``goodness'' of the fit is checked by calculating the root-mean-square difference, $rms$, between the observed and modelled magnetic field, and the relative deviation defined by $E_{rms}=rms/B_{max}$, where $B_{max}$ is the highest measured value of the magnetic field \citep[for details see][]{marubash15}.

The MC diameter is estimated as:
%
\begin{equation}
d= \frac{v\, \Delta t\, \sin\xi} {\sqrt{1-{y'_0}^2\sin^2\theta} }
\end{equation}
where $v$ is the MC speed, $\Delta t$ is the MC duration, $y'_0$ is the closest distance of the spacecraft to the MC center in the plane of the spacecraft orbit (x-y plane in SE coordinate system) normalized with respect to the MC radius, and $\xi$ is the inclination to the spacecraft-Sun line of the projection of the MC-axis  onto the plane defined by  the spacecraft-Sun line (x-axis in the SE coordinate system) and the normal to the plane of the spacecraft orbit (z-axis in the SE system). Derivation of Equation (1) is presented in https://doi.org/10.6084/m9.figshare.7599104.v1.
We also estimated $d$ applying the expression used by \cite{leitner07}, and we found no significant differences in results obtained by these two procedures.
      The fitting results, combined with the estimated values of $d$, are finally used to calculate the axial magnetic fluxes, $\Phi_{\parallel}$, and the axial electric currents, $I$ (for details see Appendix A).

Since the estimate of the flux rope extent in the measured magnetic field data is based on subjective judgement, for each event we performed three independent fittings based on three independent estimates of the beginning and end of the flux rope signature in the \emph{in situ} data. In this way we obtained three different data sets, in the following denoted as ``n'' (narrow), ``b'' (best), and ``w'' (wide), providing an assessment of the uncertainties caused by the subjective estimate of the flux rope extent.

In Table~\ref{tab1b} we display the outcome of the fitting for the b-option of the extent of flux ropes (other options will be considered only in graphs and in estimates of uncertainties of various parameters, like flux rope diameter, central magnetic field strength, axial magnetic flux, axial current, etc.). The event label is given in Column 1, which is followed by the basic MC parameters: heliocentric distance, velocity, duration, and peak magnetic field (Columns 2, 3, 4, and 5, respectively). In Columns 6\,--\,11 the results of the fitting are displayed: the latitudinal ($\theta$) and longitudinal ($\phi$) direction of the flux-rope axis (Columns 6 and 7), the impact parameter $p$ normalized with respect to the flux-rope radius (Column 8), the MC diameter $d$ (Column 9),  the magnetic field strength $B_c$ at the MC center (Column 10), and the sign of helicity $H$ (Column 11). In Columns 12 and 13 we present the $rms$ and $E_{rms}$ values, respectively.


\section{Results and Interpretation}

\subsection{Basic Concepts Considered in the Analysis}\label{S:concept}

In the absence of reconnection, the magnetic flux encircled by the erupting flux rope has to be conserved. Approximating the flux-rope by a simple line-current loop, this flux can be expressed as $\Phi_e=L_eI$, where $L_e$ is the self-inductance of the current loop and $I$ is the electric current \citep[cf.,][for the meaning of the inductance in MHD, see e.g., \citeauthor{garren94} \citeyear{garren94}, or \citeauthor{zic07} \citeyear{zic07}]{B&T62}. Since the inductance is proportional to the size of the current loop ($L\propto l$, where $l$ is the circumference of the current loop \citep[][p. 218]{jackson}, the electric current must decrease in the course of the eruption, $I\propto1/l$. Thus, also the relationship $I\propto1/R$ must be approximately valid (for the $l(R)$ relationship see Appendix B).
  Note that this is valid not only in the idealized approximation of the line-current loop, but also in the case of a flux rope of finite radius that is not constant along its axis. However, in some specific, quite
   realistic  situations, this very basic physical concept might not be applicable, e.g., when a certain set of field lines twists along a part of the flux rope, but then leaves the flux rope, becoming a set of  ``open'' field lines.

Bearing in mind that the electric current flows through a loop of finite thickness (flux rope), the total inductance is a sum of the ``external'' and ``internal'' contribution. The external inductance can be expressed in the ``semi-toroidal'' approximation \citep{chen89,garren94,zic07} as $L_e = \mu_0 l\,[\ln(8\zeta^*)-2]$, where $\mu_0$ is the permeability, $l$ is the length of the flux rope, and $\zeta^*=R/r$ is the torus aspect ratio, i.e., the ratio of the major and minor radius of the torus. The internal inductance can be expressed as $L_i = \kappa\mu_0 l$, where $\kappa$ is a constant that depends on the radial profile of the electric current density \citep[for details see][and references therein]{zic07}. Thus, in the absence of magnetic reconnection, again we get that the current has to decrease as $I\propto 1/l$. Since both $\Phi_e=L_eI$ and $\Phi_i=L_iI$ have to be conserved, also the ratio $\zeta^*=R/r$ has to be constant, i.e, the rope should expand self-similarly, $r\propto R$ \citep[for details see, e.g.,][and references therein; for a more rigorous treatment see \citeauthor{osherovich93}, \citeyear{osherovich93}]{vrs08angeo}. Thus, under these assumptions, and using the relationships presented in Appendix A and B, the following dependencies are expected: $d\propto R$, $B_c\propto R^{-2}$ , $I_{\parallel}\propto R^{-1}$, and $\Phi_{\parallel}=const.$
In this respect, let us note that statistical studies by \cite{kumar96,B&S98,liu05,wangC05,leitner07,gulisano10,gulisano12} illustrate the appropriateness of the power law presentation of the dependencies of MC parameters on the heliocentric distance.


\subsection{Results}\label{S:results}

\begin{figure}
\plotone{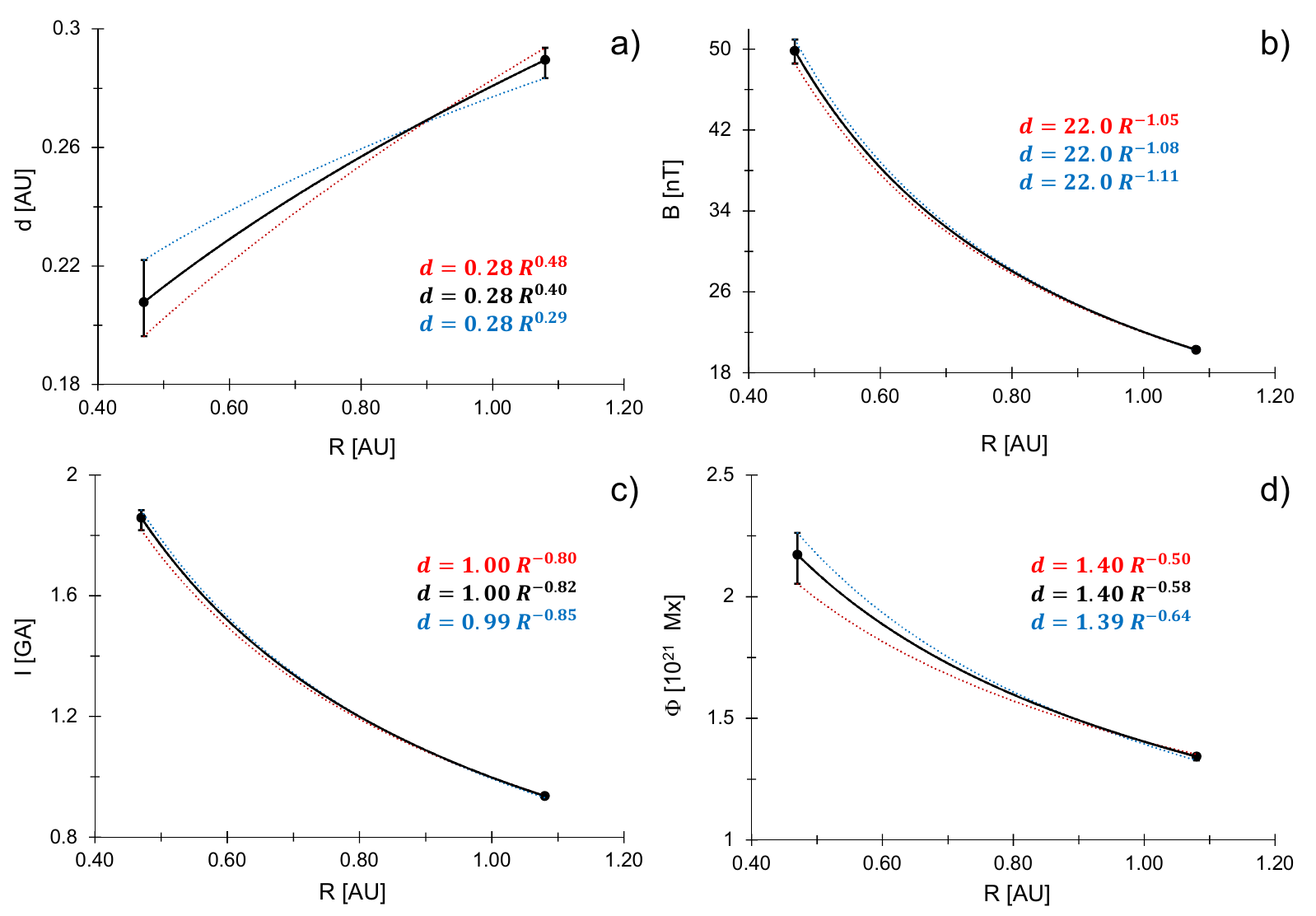}
  \caption
  {Power-law dependencies based on the results from different MC fitting procedures, shown for E8. Black marks the power-law fit through ``best" MC fit results (marked by dots), whereas blue and red line mark the extreme power-law trends obtained based on ``narrow" and ``wide" MC fits (drawn as error bars).
  \label{f1}
  }
\end{figure}

As an example of the individual-event results, we show in Figure~\ref{f1} the estimated values of the MC diameter $d$, central magnetic field $B_c$, axial electric current $I_{\parallel}$, and axial magnetic flux $\Phi_{\parallel}$ for E8. As previously mentioned, results for three fitting options (b, n, w) are presented. The data points are connected with the corresponding power-law dependencies, following the arguments explained in Section~\ref{S:concept}. To estimate the ambiguities related to the different fitting options, in addition to the power-law dependency based on the b-fit option (black line), we present in each graph also the power-law connecting the lowest value obtained for the inner spacecraft with the highest value at the outermost spacecraft (red dotted line), as well as the power-law connecting the highest value obtained for the inner spacecraft with the lowest value at the outermost spacecraft (blue dotted line). The power-law coefficients corresponding to the latter two power-law options are presented in Tables~\ref{tab2} and~\ref{tab3} as the superscripts and subscripts on the b-option value.

In Figure~\ref{f2} results for all events under study are shown together in  log-log space, where the power-law dependencies ($d=d_1 R^{\alpha_d}$, $B_c=B_{c1} R^{\alpha_B}$, $I_{\parallel}=I_1 R^{\alpha_I}$, and $\Phi_{\parallel}=\Phi_1R^{\alpha_{\Phi}}$ are represented by straight lines. Here $d_1$, $B_{c1}$,  $I_1$, and  $\Phi_1$ correspond to the MC diameter, central magnetic field, axial current, and axial magnetic flux at 1\,AU, expressed in AU, nT, GA (=$10^9$\,A), $10^{21}$ Mx, respectively. In the events where there are three data points, we applied a least-squares power-law fit.
The power-law coefficients are presented in Tables~\ref{tab2} and~\ref{tab3}, for each event individually. In addition to the power-law relations, in Tables~\ref{tab2} and~\ref{tab3}, also the linear-dependency coefficients are shown, in the analogous way as for the power-law option.

\subsubsection{Diameter and Magnetic Field Strength}\label{S:d,B}

In the case of $d(R)$ relationship shown in Figure~\ref{f2}a, the least-square power-law fit for the complete data set (dash-dotted line) reads $d=0.21R^{0.84\pm 0.29}$, with the correlation coefficient of $cc=0.53$ and the F-test confidence level of $P>99$\,\%.
However, the distribution of data points indicates that our sample consists of two statistically different subsets, one having larger dimensions, and one having significantly smaller dimensions. We checked this hypothesis by using the t-test and we found that the subsets indeed represent two statistically different populations at $P>99$\,\%. Consequently, it is worth performing independent fits for the two subsets --- the separate power law fits read $d=0.38R^{0.78\pm 0.13}$ with $cc=0.88$ (dotted line) and $d=0.11R^{0.94\pm 0.19}$ with $cc=0.85$ (dashed line), respectively. The obtained dependencies indicate that both subsets, as well as the overall fit for the whole sample show a statistical tendency broadly compatible with the self-similar expansion, i.e., that the power-law exponent is $\alpha_d\sim 1$.

On the other hand, inspecting the values of the power-law slopes $\alpha_d$ for individual events, presented in the fifth column of Table~\ref{tab2} (shown also in Figure~\ref{f3}a), one finds a large scatter, ranging from $-1.42$ (contraction) up to 2.19 (strong expansion). Note that in the following, we neglect event E2 since the two spacecraft were too close (0.13 AU; see Table~\ref{tab1} and Figure~\ref{f0}) to provide a reliable result on its heliospheric evolution. It should be noted also that the three events showing a contraction, i.e., $\alpha_d<0$ (E7, E9, E10) were measured by relatively closely-positioned spacecraft ($<0.35$ AU; see Table~\ref{tab1} and Figure~\ref{f0}). According to Table~\ref{tab2} only four events (E3, E4, E5, and E11) show $\alpha_d\simeq1$, i.e., the values that are compatible with a self-similar expansion. The large scatter of individual $\alpha_d$ values results also in a large uncertainty in the average value shown at the bottom of Table~\ref{tab2}, $\overline\alpha_d=0.38\pm1.08$. However, if the events E7, E9, and E10 (spacecraft separated by $\Delta R<0.35$\,AU)  are excluded, one finds that the remaining seven-event subsample shows $\overline\alpha_d=0.93\pm0.65$, which is consistent with self-similar expansion.


In Figure~\ref{f2}b the $B_c(R)$ dependence is presented in an analogous way as in Figure~\ref{f2}a for the $d(R)$ relationship. The least-square power-law for the complete data set (dash-dotted line) reads $B_c=16.8R^{-0.91\pm0.15}$, with a correlation coefficient of $cc=0.80$ and the F-test confidence level of $P>99$\,\%. We performed also separate fits for the two subsets like in the case of the $d(R)$ dependence, giving $B_c=13.8R^{-0.62\pm0.16}$ with $cc=0.77$ and $B_c=20.3R^{-1.34\pm0.10}$ with $cc=0.97$, for the larger and smaller MC-dimension subsets, respectively. However, in this case the subsets are not significantly different, since the F-test shows $P\lesssim95$\,\%, implying that the magnetic field strength in the considered sample does not depend significantly on the MC size. The obtained power-law slopes in any of these options are significantly different from that expected for the case of the so called isotropic self-similar expansion, meaning $B_c\propto R^{-2}$ in addition to $d\propto R$ (for a definition of ``isotropic self-similar expansion'' see e.g., \citet{demoulin09} and references therein; hereafter we will simplify it to ``self-similar expansion'').

Inspecting the values of the power-law slopes $\alpha_B$ for individual events, presented in the last column of Table~\ref{tab2} and in Figure~\ref{f3}b, one finds again a relatively large scatter of values, yet all (except the ``unreliable'' E2) showing a decrease of $B_c$ ($\alpha_B<0$). The values of $\alpha_B$ range from $-0.84$ to $-2.19$, with an average value of $\overline\alpha_B=-1.41\pm0.49$. Inspecting in detail the last column of Table~\ref{tab2}, one finds that only the events E7, E10, and E11 are compatible with a self-similar expansion ($\alpha_B\simeq -2$). In all other events one finds $\alpha_B>-2$, mostly in the range of $\alpha_B\simeq -1$, meaning that the magnetic field weakens at a significantly lower rate than expected in the case of self-similar expansion, $\alpha_B=-2$.

We end this subsection by comparing in parallel the change of the MC diameters and their magnetic field (Table~\ref{tab4}). Assuming a circular cross section of the flux rope and the magnetic flux conservation, one would expect the relationship $2\alpha_d=-\alpha_B$, since the cross-sectional area in such a case is $A\propto d^2$, i.e., $\Phi_{\parallel}\propto d^2B_c\propto R^{2\alpha_d+\alpha_B}=const.$, implying $2\alpha_d+\alpha_B=0$. The values of $2\alpha_d$ are listed in Column 2 of Table~\ref{tab4} and are compared with $\alpha_B$ (Column 3). In Column 4, the quantity $\Delta=2\alpha_d+\alpha_B$ is displayed, represented also as in the form of percentiles $\Delta_{\%} = 100\,(2\alpha_d+\alpha_B)/\alpha_B$ (Column 5). In Column 6 we present the ratio $2\alpha_d/\alpha_B$, which is expected to be $\simeq -1$ in the case of self-similar expansion.

Inspecting Table~\ref{tab4} one finds that in most events there is a large difference between $-\alpha_B$ and $2\alpha_d$, i.e. in most events the relationship $2\alpha_d/\alpha_B\simeq -1$ is not satisfied. This implies that some other effects play a significant role in the MC evolution, like e.g., magnetic reconnection that changes the magnetic flux of the rope, or a deformation of the shape of the flux-rope cross section. Yet, we note that the average value of the ratio $2\alpha_d/\alpha_B=-0.78\pm1.84$ (see the last two rows of the last column of Table~\ref{tab4}) is broadly compatible with the value expected for the case of self-similar expansion.


\subsubsection{Inferred Electric Current and Magnetic Flux}\label{S:Phi,I}

\begin{figure}
\plotone{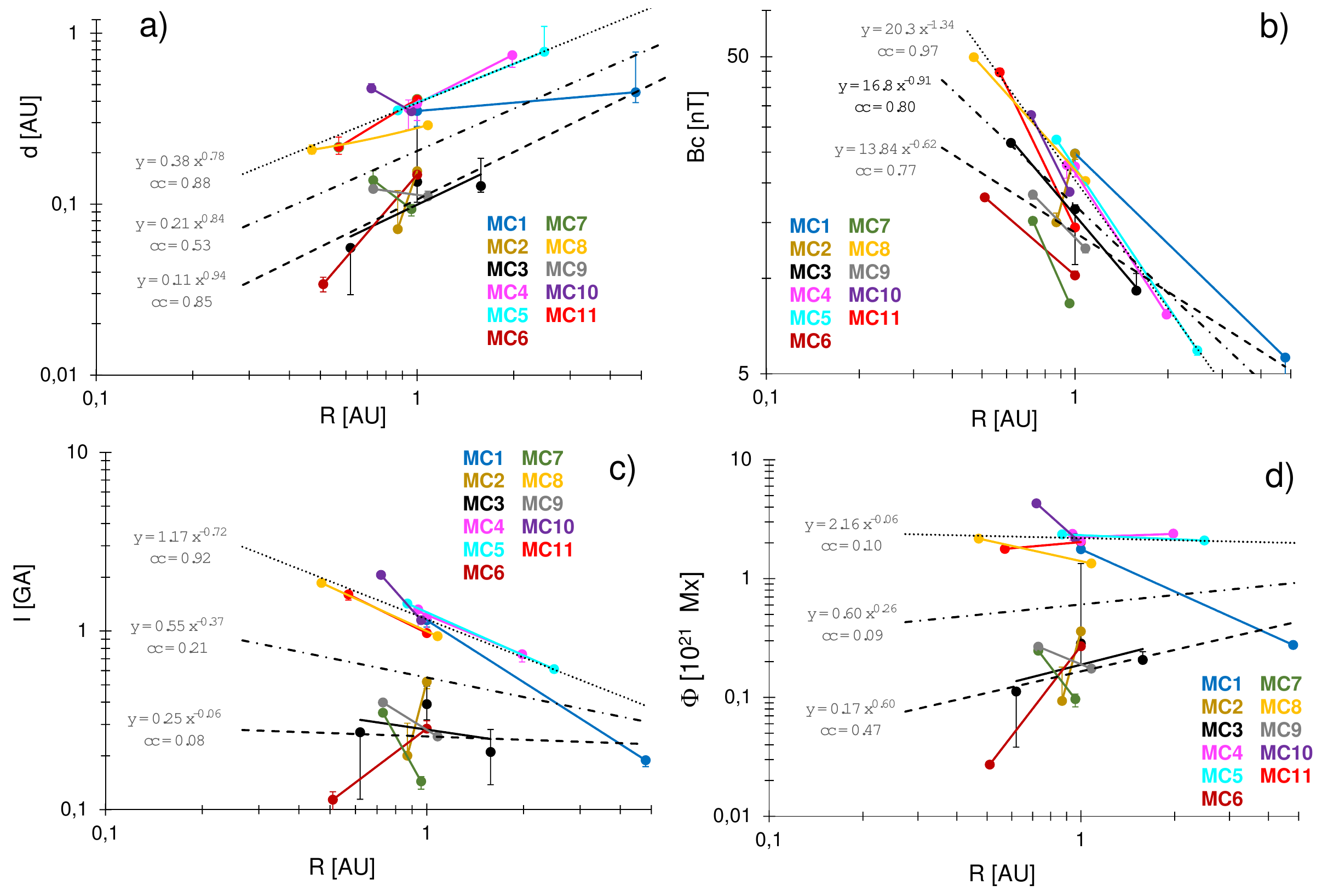}
  \caption
 {Evolution of the magnetic clouds under study presented as a function of heliocentric distance: a) MC thickness; b) central magnetic field; c) electric current; d) axial magnetic flux. Error bars are based on the ``best", ``narrow", and ``wide" MC fits.
  \label{f2}
  }
\end{figure}

In Figure~\ref{f2}c the behavior of the inferred axial electric current $I_{\parallel}(R)$ is shown for all events plotted together, including the power-law least square fits analogous to that applied in Figures~\ref{f2}a and~\ref{f2}b for the $d(R)$ and $B(R)$ dependencies. The fit for the complete data set reads $I_{\parallel}=0.55R^{-0.37\pm0.36}$, where the electric current is expressed in GA (the same applies to Table~\ref{tab3}). The corresponding correlation coefficient is $cc=0.21$. Thus, the found dependence shows a statistical tendency of decreasing $I_{\parallel}(R)$, but at a significantly lower rate than expected for self-similar expansion ($I_{\parallel}\propto R^{-1}$). The separate $I_{\parallel}(R)$ fits for the two subsets corresponding to those in the $d(R)$ dependence read $I_{\parallel}=1.17R^{-0.72\pm0.09}$ with $cc=0.92$ and $I_{\parallel}=0.25R^{-0.06\pm0.25}$ with $cc=0.08$, respectively. Thus, in a statistical sense, the subset of MCs with larger diameter show clearly a decreasing trend of the axial current, whereas smaller MCs apparently show no change of current.

\begin{figure}
\plotone{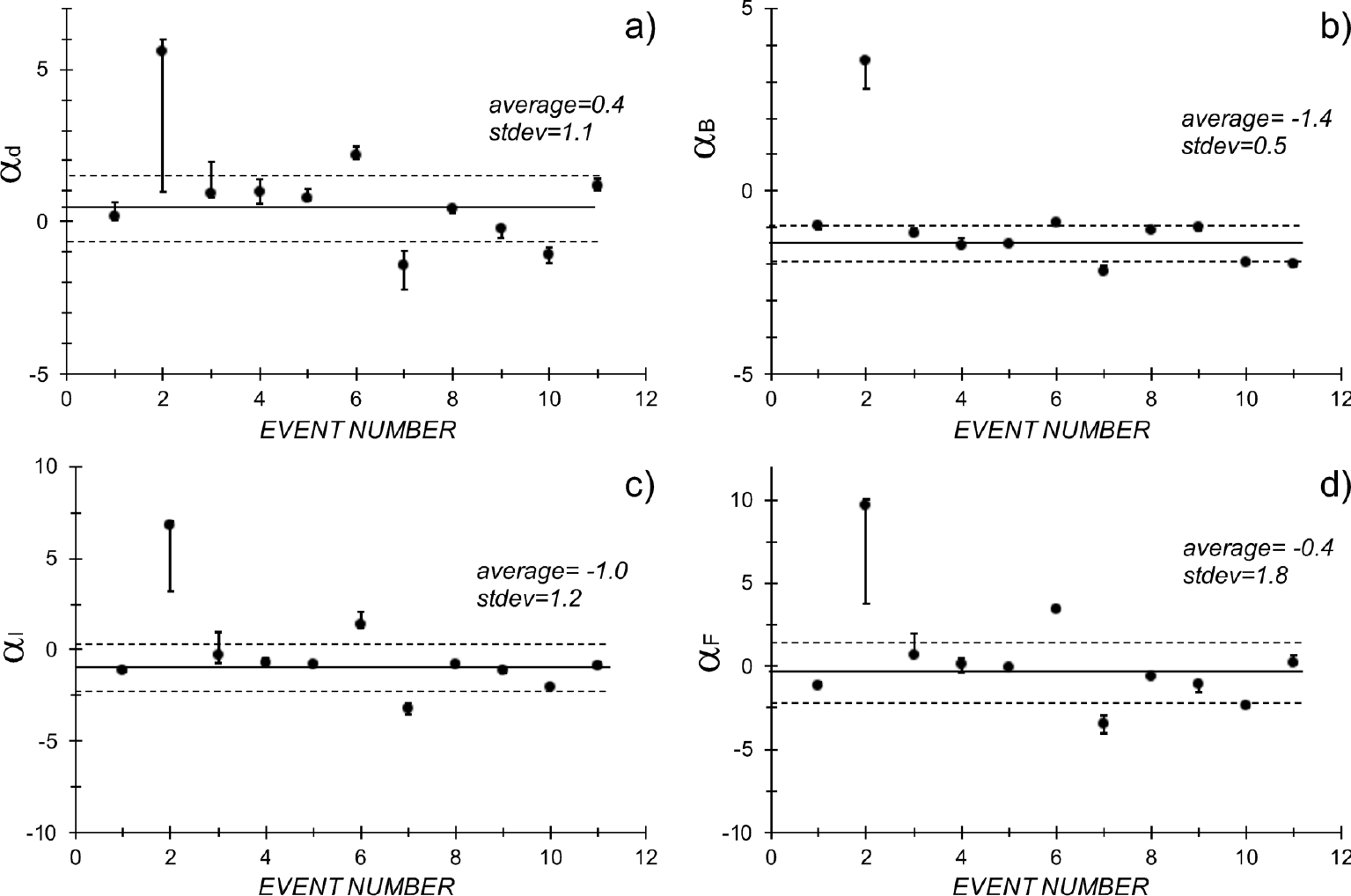}
  \caption
  {Power-law exponents for the events under study for a) MC thickness; b) central magnetic field; c) electric current; d) axial magnetic flux. Solid black lines mark mean values and  the distance of the dashed lines to the solid line equals the standard deviation (corresponding numerical values are also given in each figure). The outlier (event 2) was omitted from the calculation of the mean and the standard deviation.
 \label{f3}
  }
\end{figure}

However, inspecting the individual-event power-law exponents $\alpha_I$, displayed in Table~\ref{tab3} (shown also in Figure~\ref{f3}c), one finds that all events, except the event E6 and the ``unreliable" event E2, show a decrease of the axial current. Note that approximately half of events show $\alpha_I$ close to that expected for the self-similar expansion. The average value shown at the bottom of Table~\ref{tab3}, $\overline\alpha_I=-0.97\pm1.17$, obtained by omitting the E2 outlier (see Figure~\ref{f3}c), is compatible with that expected for the case of the self-similar expansion. If we also omit the extreme-value events E6 and E7 (see Figure~\ref{f3}c), we obtain $\overline\alpha_I=-0.98\pm0.50$, i.e., again compatible with self-similar expansion, but now with a somewhat lower standard deviation.

The behavior of the inferred axial magnetic flux $\Phi_{\parallel}(R)$ is shown in Figure~\ref{f2}d for all events plotted together, including the power-law least square fits analogous to that applied in Figure~\ref{f2}c for the $I_{\parallel}(R)$ dependence. The fit for complete data set reads $\Phi_{\parallel}=0.60R^{+0.26\pm0.59}$, where the magnetic flux is expressed in units of $10^{21}$ Mx (the same applies to Table~\ref{tab3}). The corresponding correlation coefficient is only $cc=0.09$. Thus, bearing in mind a large uncertainty of the power-law exponent and very low correlation coefficient,
  as well as the fact that the flux rope interval chosen at one location may not match entirely the interval chosen at the other,
   the axial magnetic flux can be considered as constant.
The separate $\Phi_{\parallel}(R)$ fits for the same two subsets read $\Phi_{\parallel}=2.16R^{-0.06\pm0.19}$ with $cc=0.10$ and $\Phi_{\parallel}=0.17R^{+0.60\pm0.36}$ with $cc=0.47$, i.e., the former one is compatible with
$\Phi_{\parallel}(R)=const.$, whereas the latter one indicates a weak, yet statistically insignificant, increasing trend of $\Phi_{\parallel}$.

On the other hand, the average value $\overline\alpha_{\Phi}=-0.43\pm1.84$, presented at the bottom row of Table~\ref{tab3}, is broadly compatible with $\Phi_{\parallel}(R)=const$. If we exclude E2 and the extreme-value events E6 and E7 (see Figure~\ref{f3}c), we find $\overline\alpha_{\Phi}=-0.54\pm0.97$, again being broadly compatible with $\Phi_{\parallel}(R)=const.$

\begin{figure}
\plotone{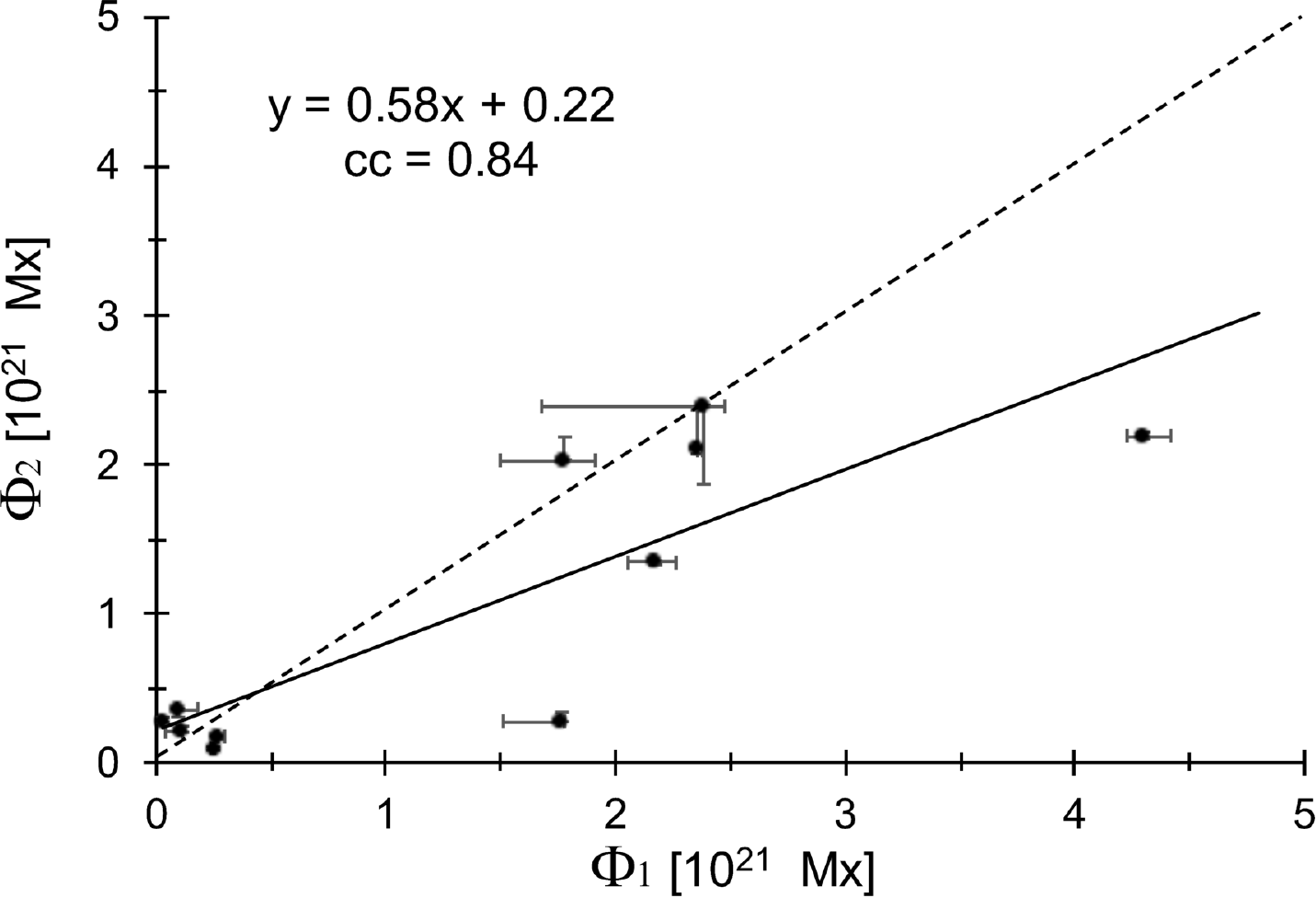}
  \caption
  {Change of the inferred MC magnetic flux: $\Phi_1$ and $\Phi_2$ represent the axial magnetic flux, expressed in units of $10^{21}$\,Mx, at the first and the last spacecraft measurement for each event. The dashed line represents the $\Phi=const.$ case. The solid line shows the least squares linear fit to the data points.
  \label{f4}
  }
\end{figure}

In Figure~\ref{f4} the effect of decreasing axial magnetic flux is emphasized by showing for each event the flux inferred from the measurements at the farthermost spacecraft, $\Phi_2$, versus the flux obtained for the spacecraft closest to the Sun, $\Phi_1$. The graph shows that data points tend to be below the $\Phi_2=\Phi_1$ line, i.e., in most of events the value of $\Phi_2$ is smaller than $\Phi_1$. The linear least square fit to the data points reads $\Phi_2=(0.58\pm0.16)\Phi_1-(0.22\pm0.5)$, with a correlation coefficient of $cc=0.84$. A linear fit fixed at the origin, gives $\Phi_2=0.67\,\Phi_1$, with $cc=0.82$. Note that the correlation is dominated by events with large values of $\Phi_1$ ($>10^{21}$\,Mx), where all events show either $\Phi_2<\Phi_1$ (E1, E8, E10) or $\Phi_2\sim \Phi_1$ (E4, E5, E11). For the six events with large $\Phi_1$, the average relative decrease $\Delta\Phi/\Phi_1$ amounts to $\sim30$\,\%, consistent with the mentioned fit $\Phi_2=0.67\,\Phi_1$.



\section{Discussion}

The relationships expected for self-similar expansion of the cylindrical force-free flux rope (Section~\ref{S:concept}) are not fully consistent with the observations presented in Section~\ref{S:results}. Even taking into account only the measurements where the spacecraft were radially separated by more than 0.4 AU, where the statistical trend is compatible with the self-similar expansion form $d\propto R$,
 individual events show a great variety of behaviors, from very weak to very strong expansion (Table~\ref{tab2}). According to Table~\ref{tab2} only four events (E3, E4, E5, and E11) show $\alpha_d\simeq1$.

 The behavior of the magnetic field strength shows an even more significant deviation from that expected for the self-similar expansion, even in the statistical sense. The overall fit through whole data set (Figure~\ref{f2}b) shows that the rate at which $B_c$ decreases is characterized by $\alpha_B\sim -1$, which is significantly lower than that for self-similar expansion ($\alpha_B=-2$). Accordingly, most of the events individually show a  similar behavior, resulting in a mean value of $\overline\alpha_B\sim -1.4$. Table~\ref{tab2} shows that only three events (E7, E10, and E11) are compatible with a self-similar expansion ($\alpha_B\simeq -2$). Finally, Figure~\ref{f4} indicates that in the statistical sense, the axial magnetic flux decreases with distance.
 Thus, the analyzed set of events might indicate that there is a significant reconnection reducing the MC size and the magnetic flux by ``peeling-off'' the outer layers of the rope \citep[][]{dasso06,dasso07,gosling07,mostl08,ruffenach12}, or more likely, that the assumption of a circular cross section is, at least in a fraction of events, not valid.
   In this respect, it should be emphasized that the imperfection of existing models is the leading cause of uncertainty in reaching a definitive conclusions; note that the flux erosion as inferred here and in previous papers, is based on usage of very simplified models, so the validity of such results remains questionable.

\subsection{Nonuniform Flux-Rope Expansion}\label{S:asym}

\begin{figure}
\epsscale{0.50}
\plotone{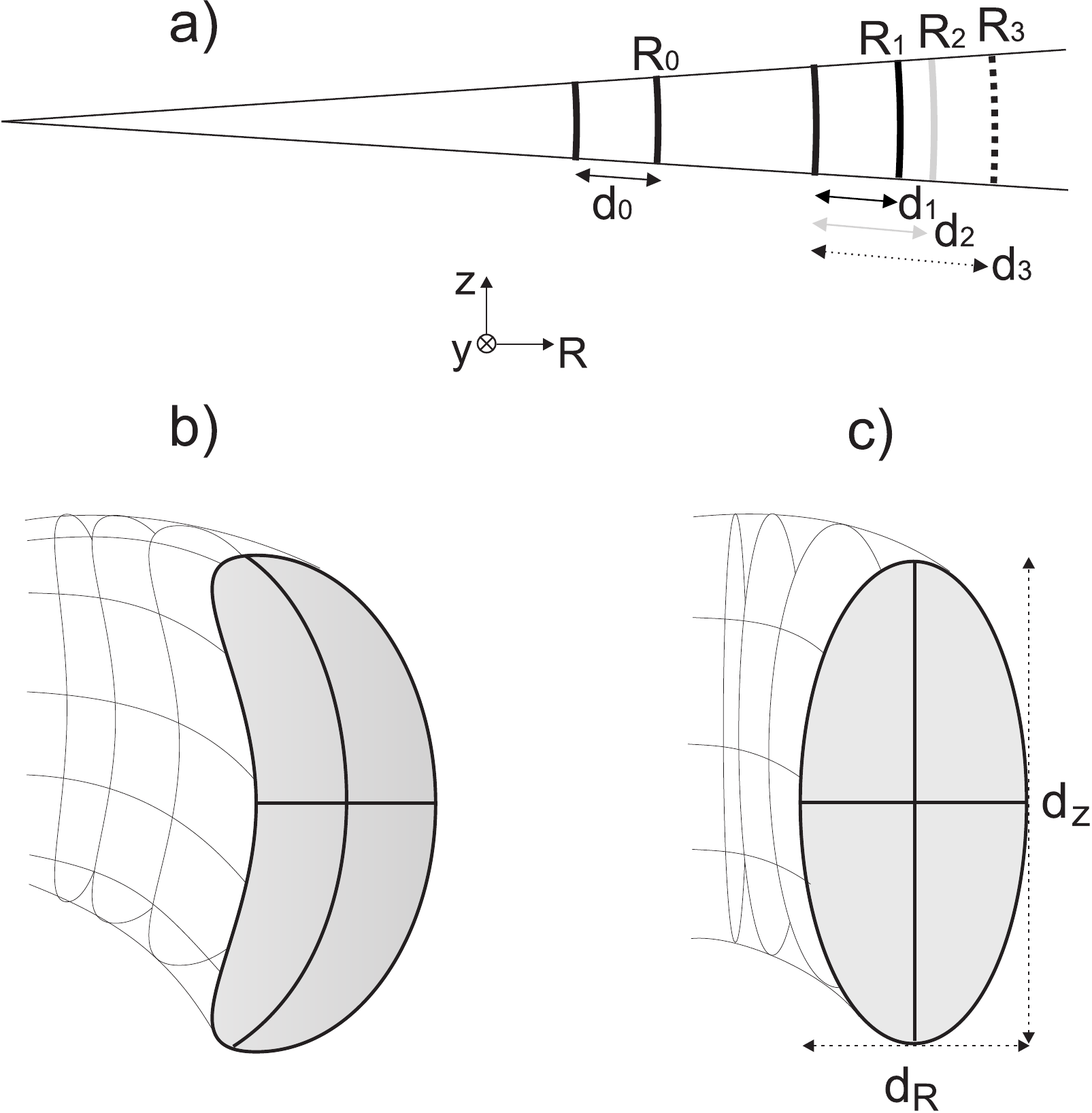}
  \caption
  {
Nonuniform flux-rope expansion. a) Schematic sketch of different variants of the radial expansion of an element of the flux-rope; b) Presumed form of the flux-rope expansion; c) Approximation in terms of an elliptical flux-rope cross-section. For details see the main text.
  \label{f5}
  }
\end{figure}

Let us first consider the effects of a ``nonuniform'' expansion, schematically drawn in Figure~\ref{f5}a. The axial magnetic field is oriented in $y$-direction. In $z$-direction the considered element expands in a self-similar manner, whereas for the radial expansion three options are taken into account. Initially, the element has a thickness $d_0$, with the frontal edge set at a heliocentric distance $R_0$. In the course of time, the element propagates to a larger distance, $R_i$, and attains a thickness $d_i$ ($i=1$, 2, 3).

The expansion depicted by the bold-black frontal edge at the distance $R_1$ represents the option where the element does not change its thickness, i.e., $d_1=d_0$, so the cross-section area behaves as $A\propto R$. Note that in such a case there should be no velocity gradients within a MC ($v_{front}=v_{rear}$), i.e., the MC speed is fully adjusted to the ambient solar wind speed.
Bearing in mind the flux conservation, the magnetic field in the element would decrease as $B\propto R^{-1}$, i.e., much slower than in the case of the self-similar expansion where it is characterized by $B\propto R^{-2}$. Such a behavior is found (see Table~\ref{tab2}) in events E1, E3, E6, E8, and E9, where all but E3 and E6 show only a very weak or no expansion.

The self-similar expansion is depicted in Figure~\ref{f5}a by the bold-dotted line at $R=R_3$, where $d_3/d_0=R_3/R_0$. Here, the expansion should be characterized by $\alpha_d=1$ and $\alpha_B=-2$. The only event that satisfies both conditions is E11. The events E7 and E10 show $\alpha_B\simeq -2$, however $\alpha_d$ in these events corresponds to the MC contraction (yet, as mentioned in Section~\ref{S:d,B}, the values of $\alpha_d$ are based on measurements by two relatively closely located spacecraft). In the events E3 and E4 we found $\alpha_d\simeq 1$, however the decrease of $B_c$ is much weaker than in the case of self-similar expansion.

The expansion denoted by the bold-gray line at $R=R_2$ represents an intermediate case, where $d_0<d_2<d_3$. In this option, the field decreases as $B\propto R^{-k}$, where $1<k<2$, just like it was found (in the statistical sense) from the observations in Section~\ref{S:d,B}. Such behavior is found in E4 and E5.

Based on numerical simulations of the flux-rope propagation it can be expected that the expanding flux rope has the outward-convex shape depicted in Figure~\ref{f5}b (``pancaking effect'') due to pressure gradients and/or the MHD ``aerodynamic'' drag \citep[see, e.g.,][]{cargill94,cargill96,cargill99,riley04,owens06}. In the interest of simplicity, let us approximate such a structure by an elliptical cross-section, as drawn in Figure~\ref{f5}c \citep[for the magnetic field configuration in such a rope see, e.g.,][]{vandas04,riley04fit}. In particular, we assume that the expansion in the direction perpendicular to the plane of the flux rope axis, which is set in the $R$-$y$ plane is proportional to $R$. This means, that for the thickness in $z$-direction $d_z\propto R$ is valid, i.e., $d_z/R=const.$, meaning that the major axis of the cross-sectional ellipse expands in a self-similar manner. For the thickness in the $R$-direction we allow that the ratio $d_R/R\equiv d/R$ is a function of distance, in particular, $d\propto R^{\alpha_d}$. In that case the cross-sectional area behaves as $A=d R\theta \pi/4$, where $\theta=const.$ represents the MC heliocentric angular width in the $z$-direction. Under the described approximations and in the absence of reconnection, the magnetic field should decrease as $B\propto R^{-(1+\alpha_d)}$, i.e., the relation $\alpha_B =-(1+\alpha_d)$ should be valid.

In Column 7 of Table~\ref{tab4} the values of $1+\alpha_d$ are listed, and in Column 8 we display the ratio $-(1+\alpha_d)/\alpha_B$. One finds that in the E1, E4, E5, E8, E9, and E11, $\alpha_B$ differs from $1+\alpha_d$ within 30\,\%. Considering the measurement uncertainties, this is consistent with the assumption that the expansion in $z$-direction is confined to $\theta\simeq const.$, i.e., the measurements could be interpreted by nonuniform expansion without including some additional effects.

On the other hand, in the remaining four events (again excluding E2), E3, E6, E7, and E10,
the difference of $1+\alpha_d$ and $\alpha_B$ is much larger and the magnetic field decrease cannot be explained solely by the $\theta=const.$ expansion. Consequently, such a decrease of the magnetic field indicates that either the magnetic flux was significantly reduced by reconnection, or these MCs significantly ``over-expanded'' in $z$-direction, i.e., $\Delta\theta/\Delta R >0$. Note that the latter effect is found in numerical simulations presented by \citet{riley04}. However, note that E7 and E10 were measured by two relatively closely located spacecraft, so here the results are quite uncertain, and the difference might be only due to errors in measurements. Also note that E3, E6, and E7 were events of low magnetic flux, so reconnection can play a more important role than in the case of large-flux events.


\subsection{Magnetic Reconnection}

\begin{figure}
\epsscale{0.70}
\plotone{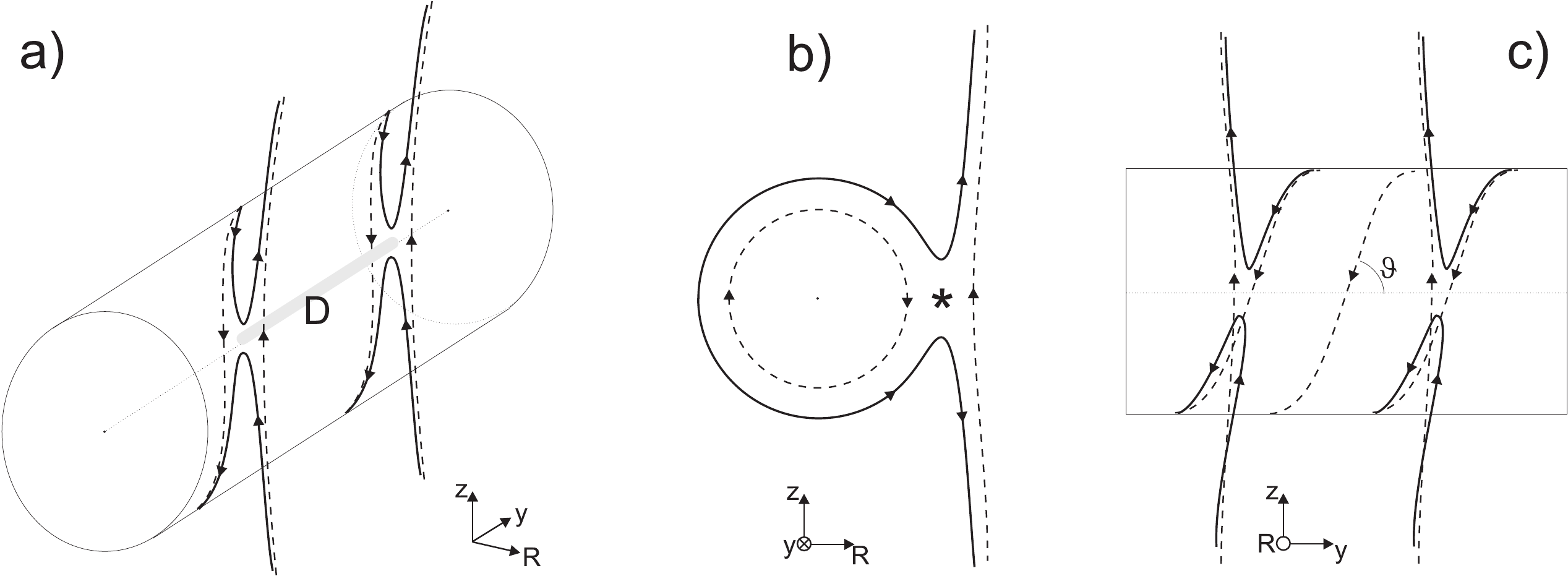}
  \caption
  {Sketch of flux-rope helical magnetic field reconnecting with the external field: a) 3-D presentation; b) View along the flux-rope axis; c) View perpendicular to the flux-rope axis. The pre-reconnection field lines are drawn dashed, the reconnected ones are depicted by full lines, and the diffusion region (X-type neutral line) is indicated by thick-gray line in a) and by asterisk in b).
  \label{f6}
  }
\end{figure}

In Figure~\ref{f6}, reconnection of the helical field of the rope and the external field is sketched. In the following it is assumed that the reconnection takes place at the leading edge of the MC over the length $D$, and we take approximately $D\approx R\,\omega$, where $\omega$ is the angular extent of the flux rope in the plane of its axis ($R$-$y$ plane in Figures~\ref{f5} and~\ref{f6}). A local reconnection rate is determined by the product $v_iB_i$, where $v_i$ is the inflow speed of the magnetic field $B_i$ into the diffusion region (note that value of $v_iB_i$ has to be equal on the opposite sides of the diffusion region).
Thus, the rate at which the magnetic flux is reconnected over the length $D$ equals to $\Delta\Phi/\Delta t=D\,v_iB_i$, where the values of $D$, $v_i$, and $B_i$ depend on the heliocentric distance.

Under these assumptions, the total reconnected flux in the time interval from $t_a$ to $t_b$ amounts to:
\begin{equation}
   \Delta\Phi=\int_{t_a}^{t_b} DB_iv_i \,{\rm d}t\,,
\end{equation}
which can be expressed by integrating over the corresponding heliocentric distance range as:
\begin{equation}
   \Delta\Phi=\int_{R_a}^{R_b} \frac{DB_iv_i}{\overline V_{\rm MC}} \,~{\rm d}R\,,
\end{equation}
where we used d$R$\,=\,$\overline V_{\rm MC}$\,d$t$ for the MC propagating at an average speed $\overline V_{\rm MC}$. In the following, we take that the ambient interplanetary magnetic field decreases approximately as $R^{-2}$ \citep{vrs04bsplitIII}, i.e., we take $B_i\propto R^{-2}$.
Assuming that the solar wind expands at a constant speed and that the expansion is approximately isotropic, one finds that the density varies as $\rho\propto R^{-2}$ \citep[e.g.,][]{vrs13dbm}, so the Alfv\'en speed should decrease roughly as $v_A=B_i/(\mu \rho)^{1/2}\propto R^{-1}$. Thus, $B_i$, $n$, and $v_A$ can be expressed as:
\begin{equation}
   B_i(R) = B_1 R^{-2}~~~{\rm and}~~v_A(R) = v_{A1} R^{-1}\,,
\end{equation}
where subscript 1 stands for the value at 1\,AU, and $R$ is expressed in AU. The reconnection inflow speed is a small fraction of the local Alfv\'en speed $v_i(R)=\kappa\,v_A(R)$, with $\kappa=0.01$\,--\,0.1 \citep[e.g.,][and references therein]{priest}. In such a case, the total flux that is reconnected in the course of the MC propagation from $R_a$ to $R_b$ equals to:
\begin{equation}
   \Delta\Phi=C_r\int_{R_a}^{R_b} R^{-2}\,{\rm d}R\,,
\end{equation}
where $R$ is expressed in AU, and $C_r$ stands for:
\begin{equation}
   C_r= \kappa\, \omega \,\frac{B_1\, v_{A1}}{\overline V_{\rm MC}}\, \times (1.5\times 10^{11})^2\,.
\end{equation}
Here, the numerical factor $1.5\times10^{11}$\,m represents 1\,AU, and if the ambient field $B_1$ is expressed in T, Equation (5) provides the results in Wb, which corresponds to $10^8$\,Mx.
After integrating Equation (5), the total flux reconnected while the MC travels from $R_a$ to $R_b$ reads:
\begin{equation}
   \Delta\Phi=C_r \left(\frac{1}{R_a} - \frac{1}{R_b}\right)\,.
\end{equation}
Substituting the values $\omega=1$\,rad, $\kappa=0.1$, $B_1=5$\,nT, $v_{A1}=50$\,km\,s$^{-1}$, and $\overline V_{\rm MC}=500$\,km\,s$^{-1}$, one finds that in the case of fully antiparallel fields of MC and ambient solar wind, the flux reconnected over a distance range from $R_a=1$ to $R_b=2$ equals to $\Delta\Phi=5.6\times10^{11}$\,Wb\,$=5.6\times10^{19}$\,Mx. This is an order of magnitude lower than estimated in Section~\ref{S:Phi,I}, and is consistent with the results presented by \cite{gosling05}. On the other hand, closer to the Sun, the reconnected flux could be much larger, as inferred also by \cite{ruffenach12}. For example, for $R_a=0.1$ and $R_b=1$ one gets $10^{21}$\,Mx.

In Table~\ref{tab5} the observed change of the magnetic flux from the first to the last spacecraft measurement
(only the decreasing-$\Phi$ events are shown)
 is compared with the reconnection-related change calculated using Equation~(11). The ratio of these two values, given in Column 7, shows that the observed flux change is larger than that presumably caused by reconnection. We note that for the calculation of the reconnected flux we used $\kappa=0.1$, which is an upper limit for the reconnection rate, so the calculated values represent an upper limit, particularly considering the most favorable case of fully antiparallel field of MC and ambient solar wind. The only event where the observed
and calculated changes of the flux coincide is E9, where the initial magnetic flux is very low. We also note that the observed values of $\Delta\Phi$ do not show any statistically significant dependence on the heliospheric distance that is predicted by Equation (7). These findings imply that reconnection alone cannot explain the inferred magnetic flux changes, since its effect is more than one order of magnitude weaker than required.

Finally, we stress that Equation (7), due to the dependence $C_r\propto \overline V_{MC}$, indicates that in the case of fast MCs the effect of reconnection is expected to be much weaker than in the case of slow MCs, which is related to shorter time available for reconnection. For example, taking $\overline V_{MC}\approx 1000$\,km\,s$^{-1}$, and using the set of parameter values $\omega=1$\,rad, $\kappa=0.1$, $B_1=5$\,nT, and $v_{A1}=50$\,km\,s$^{-1}$, one finds that from $R_a=0.5$ to $R_b=1$, less than $0.06\times10^{21}$\,Mx is expected to be reconnected. Since fast MCs are usually characterized by strong magnetic field, implying also a large magnetic flux, in such events the reconnection-related relative decrease of the magnetic flux is probably negligible. On the other hand, in the case of slow MCs, the reconnection can be a significant factor. Taking the same parameters as in the previous example, only now substituting  $\overline V_{MC}\approx 300$\,km\,s$^{-1}$, one finds that a flux of $\approx2\times10^{21}$\,Mx could be reconnected, implying that low-flux MCs may be entirely ``melted'' into the background solar wind before reaching the Earth.

Finally, let us also note that reconnection affects not only the magnetic flux of MCs, but also their diameter, since ``peeling-off'' the outer layers of the flux rope should lead to lowering the expansion rate of MCs. This might at least partly explain the expansion rate characterized by $\alpha_d<1$.


\subsection{Comparison with Previous Studies, and Accuracy of the Results}

In the following, the results presented in Section~\ref{S:results} are compared with the results of previous studies on the heliospheric evolution of the MC size and magnetic field. First we give an overview of the statistical aspects, and then we focus on the evolution of individual events.

The slope of the ``overall'' power-law fit $d(R)$ to all data shown in Figure~\ref{f2}a reads $\alpha_d=0.84\pm0.29$. This value falls into the range found in the statistical studies by \cite{kumar96}, \cite{B&S98}, \cite{liu05}, \cite{wangC05}, \cite{leitner07}, \cite{gulisano10}, and \cite{gulisano12}, where $\alpha_d$ is found in the range from $0.49\pm0.26$ \citep[][]{gulisano12} to $0.97\pm0.1$ \citep[][]{kumar96}.
On the other hand, the mean value of the slopes obtained for individual events, $\alpha_d=0.38\pm1.08$ (Table~\ref{tab2}), is considerably lower than most of previously reported values. However, omitting the events E7, E9, and E10 (spacecraft separated by $\Delta R<0.35$\,AU, showing MC contraction), the remaining seven-event subsample gives $\overline\alpha_d=0.93\pm0.65$, which is consistent with self-similar expansion.

 In this respect, it should be noted that all of the previously mentioned studies show large variety of $\alpha_d$ values, the differences generally being larger than the reported error estimates. Moreover, as demonstrated in Section~\ref{S:d,B}, the statistical results based on a sample of single spacecraft measurements could be misleading since the outcome depends on the distribution of data points over a given distance range, particularly bearing in mind the effect of weakening of the expansion with increasing heliocentric distance \citep[e.g.,][]{osherovich93,liu05,wangC05,leitner07,gulisano12}.

In any case, we emphasize that most of the mentioned studies show the $d(R)$ dependence with $\alpha_d<1$, i.e., reveal a deviation from self-similar expansion. We also note that the back-extrapolation of both the power-law and the linear $d(R)$ relationships lead to far too large MC sizes in the solar vicinity. From this one can infer that the MC expansion is much more pronounced at small heliocentric distances, consistent with the results presented by \citet{leitner07} and \citet{wangC05}.

When the evolution of individual events is considered, our results could be compared with only a few case studies. \cite{B&S98}  presented as example a MC observed by Helios 1 at 0.9 AU and by Voyager 1/2 at 2.6 AU, roughly doubling its thickness over this distance range. This would correspond to $\alpha_d\sim0.65$, which is comparable to our events E5 and E8.
A similar expansion rate was found by \cite{savani09}, who analyzed the Heliospheric Imager remote observations of one circularly-shaped CME.

\cite{chinchilla12} performed a very detailed study of the evolution of a MC recorded by the MES at $R\sim0.5$\,AU and by the Wind spacecraft at 1\,AU. They applied several magnetic reconstruction techniques and various assumptions on the MC boundaries, to demonstrate how much the results depend on the methodology applied. The differences turned out to be very large in determining the MC diameter, resulting in very different expansion rates --- from the data presented in the paper one finds cases from shrinking with $\alpha_d=-0.8$, up to expansion with $\alpha_d=1.1$. The expansion, $d(R)$,  was also traced from the HI data in the range $R<0.6$\,AU, which helped to resolve this ambiguity, giving the expansion rate in the direction of motion consistent with that found statistically by \cite{B&S98}. Note that the results displayed in Table~\ref{tab2} show a similar spread of $\alpha_d$, ranging from $-1.42$ to $+2.19$.


From the results presented in Table~\ref{tab2} one finds $B(R)$ power-law slopes ranging from $\alpha_B=-0.84$ to $-2.19$.
This is consistent with the power-law slopes found by \cite{kumar96}, \cite{liu05}, \cite{wangC05}, \cite{leitner07}, \cite{gulisano10}, and \cite{gulisano12}, that range from $\alpha_B=-0.88\pm0.22$ \citep[][]{leitner07} to $\alpha_B=-1.85\pm0.11$ \citep[][]{gulisano10}, again showing a large scatter of values.

 The ``overall'' power-law fit to all data points shown in Figure~\ref{f2}b has a slope of $\alpha_B=-0.91\pm0.15$, whereas the mean value of the slopes obtained for individual events gives  $\alpha_B=-1.41\pm0.49$ (Table~\ref{tab2}). The former value is close to that found by \cite{leitner07}, whereas the latter value is close to $\alpha_B=-1.3$ obtained by \cite{du07}. Inspecting the dependence of the $B(R)$ slopes on the heliocentric distance in the mentioned papers, one finds that the decrease of the magnetic field is faster closer to the Sun than at large distances ($\alpha_B\sim -1.8$ in the range 0.3\,--\,1 AU, versus $\alpha_B\sim -0.9$ to $\sim -1.4$ in the range 1.4\,--\,5.4 AU), which is consistent with weakening of the expansion with increasing distance.



\cite{osherovich93} studied theoretically a self-similar expansion of MCs and compared the results with the \emph{in situ} measurements of a MC recorded by Helios 2 at 1\,AU and Voyager 2 at 2\,AU. It was shown that over this distance range the central magnetic field decreased by a factor of four, corresponding to $\alpha_B=-2$, i.e., being consistent with the self-similar expansion.

\cite{chinchilla13}, studied an ICME identified in the MES data (located at $R\sim0.56$\,AU), which was also recorded at the STB spacecraft that was aligned with MES at the distance of $R=1.08$\,AU. Although the evolution of the internal magnetic field structure was not analyzed in detail, from the data presented in the paper it can be inferred that the thickness of the MC contained in the ICME was increasing at the rate $\alpha_d\sim0.8$. The measured (i.e., not reconstructed) magnetic field was decreasing at the rate $\alpha_B\sim -0.9$, thus also showing a behavior that is inconsistent with self-similar expansion.

\cite{good15} analyzed a MC observed by MES located at 0.44 AU and later by the STB spacecraft at 1.09 AU in early November 2011. The applied force-free fitting showed that the MC size was increasing at the rate $\alpha_d\sim0.91$, whereas the magnetic field decrease was characterized by $\alpha_B\sim -1.84$. The analysis showed that the axial magnetic flux was conserved, i.e. no significant erosion took place between 0.44 and 1.09 AU.

\cite{du07} studied the evolution of the magnetic flux and helicity of the MC recorded by ACE at Earth on 4-6 March 1998 and later by Ulysses at 5.4 AU. Applying the Grad-Shafranov reconstruction technique, they found that the inferred value of the axial flux decreased by an order of magnitude. From the presented values of the peak axial magnetic field component one finds that it decreased at a rate of $\alpha_B\sim -1.3$, the size increase was characterized by $\alpha_d\sim0.4$, whereas the axial-flux decrease is characterized by $\alpha_{\Phi}\sim -1.1$ to $-1.5$. These rates are in the range of the values presented in Table~\ref{tab2}.

\cite{mulligan01grl} studied the evolution of the MC associated with the ``Bastille-day flare''
employing the \emph{in situ} data from the ACE and NEAR spacecraft that were located at $\sim1$ and $\sim1.78$\,AU. They found $\alpha_B=-1.4$, which is very close to the mean value displayed in Table~\ref{tab2} ($\alpha_B=-1.41$).
It was also inferred that the magnetic flux increases at a rate of $\alpha_{\Phi}=0.63$.

The discussion presented in the previous paragraphs shows that the conclusions on the evolution of MCs can be quite ambiguous since the empirical results depend on a number of factors. For example, the outcome of the magnetic field reconstruction strongly depends on the level of complexity of the true magnetic structure of a given MC and the trajectory of the spacecraft. Furthermore, different reconstruction methods give different results \citep[see, e.g.,][]{dasso06}. Finally, the outcome to a certain degree depends on the interpretation of the measurements, e.g., the estimate of the MC boundaries, which is subjective and often differs from author to author.

Regarding the errors and reliability of the results, it should be noted that it is quite difficult to estimate the accuracy of the results in the case when the analysis is based on only two spacecraft.
In this respect, our study provides a certain insight into the errors since in the events E3 and E4 measurements from three different \emph{in situ} observatories are available (see Table~\ref{tab1} and Figure~\ref{f0}). In E4 the first two spacecraft were located at similar heliocentric distances, 0.94 and 1 AU (Figure~\ref{f0}), and the reconstruction resulted in very similar outcome (see measurements 4a and 4b in Table~\ref{tab1b} and check E4 Figure~\ref{f2}).

On the other hand, it should be noted that the situation was quite different in the case of E2, where measurements were also performed at two relatively closely spaced spacecraft (0.87 and 1 AU). In this event the outcome for the two spacecraft were quite different, particularly in the case of the diameter $d$ and the reconstructed central magnetic field $B_c$ (see measurements 2a and 2b in Table~\ref{tab1b} and check E2 in Figure~\ref{f2}). Similarly, in E3, where the three spacecraft were located at $R$\,=\,0.62, 1, and 1.58 AU, the evolution of the estimated MC velocities $v(R)$ shows a considerable scatter ($v(R)$\,=\,289, 462, and 376 km\,s$^{-1}$, respectively), which results also in a considerable scatter in the diameter evolution ($d(R)$\,=\,0.056, 1.35, and 0.128 AU, respectively; see measurements 3a, 3b, and 3c in Table~\ref{tab1b} and check E3 in Figure~\ref{f2}). On the other hand, the values of $B_c$ show a smooth decay with $\alpha_B=-1.15$. However, due to the scatter in the $d(R)$ dependency, the inferred dependencies $I_{\parallel}(R)$ and $\Phi_{\parallel}(R)$ also show a significant scatter.

\section{Conclusions}

We presented a study of the evolution of eleven MCs based on the \emph{in situ} measurements by at least two radially aligned spacecraft. The analysis has shown that reliable results can be obtained only if the spacecraft separation is $\gtrsim 0.5$ AU. It is also shown that  there is a large difference between the behavior of individual events and the overall statistical trends. Thus, overall fits, like those presented by \cite{kumar96}, \cite{B&S98}, \cite{liu05}, \cite{wangC05}, \cite{leitner07}, \cite{gulisano10,gulisano12} should be taken with some caution, since they can lead to wrong physical interpretations of individual events.
Bearing in mind these two facts, the results of our study can be summarized as follows.

\begin{itemize}
\item In the statistical sense MCs in the sample show an expansion compatible with self-similar expansion ($d\propto R$). However, individual events show a large scatter of expansion rates, ranging from very weak to very strong expansion; only four events show an expansion rate compatible with self-similar expansion. The results indicate that the expansion has to be much stronger when MCs are still close to the Sun.
\item The magnetic field shows a large deviation from the behavior expected for the case of a self-similar expansion. In the statistical sense, as well as in most of individual events, the inferred magnetic field decreases much slower than expected. Only three events show a behavior compatible with self-similar expansion.
\item The presented analysis indicates that there is also a discrepancy between the magnetic field decrease and the increase of the MC size, suggesting that magnetic reconnection and the ``pancaking effect'' might play a significant role in the MC evolution. However, bearing in mind the usage of very simplified models, as well as the fact that the reconstruction of the magnetic configuration is based on single-point time series, this indication has to be taken with caution.
\item Individually, about half of the events show the decay of the  electric current as expected in the case of self-similar expansion, which is also reflected in the mean value of the decay rate.
\item In the statistical sense, the inferred axial magnetic flux is broadly consistent with staying constant during the MC evolution. However, events characterized by large magnetic flux show a clear tendency of decreasing flux.
\end{itemize}

The presented analysis shows some significant deviations from the behavior expected for self-similar evolution of MCs. In some events the diameter increases at a rate much lower than $d\propto R$, which might be explained by gradual adjustment of the MC dynamics to the ambient solar-wind flow. Generally, there is a tendency that ejections that are faster than the solar wind decelerate, whereas those that are slower accelerate during their interplanetary propagation \citep[e.g.,][]{gopal00}, as a consequence of the ``magnetohydrodynamic drag'' \citep[e.g.,][and references therein]{cargill04,vrs08dij,vrs13dbm}. Due to the same effect one would expect that the broadening of the MC body gradually weakens, until eventually all elements attain the speed of the ambient solar wind. Since the overall solar-wind flow is characterized by a constant velocity, this implies that all elements of the MC should attain the same speed. Consequently, there should be no change of the MC diameter at large heliocentric distances. It should be noted that the absence of velocity gradients, and the related disappearance of the frontal sheath region, would make the identification of ICMEs in the \emph{in situ} data considerably more difficult.

Qualitatively, the reduced expansion rate is consistent with a slower decrease of the MC magnetic field than is expected in the case of self-similar expansion. Yet, there is a considerable discrepancy in quantitative terms in several events, where the decay of the magnetic field is not consistent with the expansion rates.

An apparently plausible way to explain the mentioned discrepancies, is to presume that they are due to magnetic reconnection occurring between the internal MC field and the ambient interplanetary field
\citep[][]{dasso06,dasso07,gosling07,mostl08,ruffenach12}.
The effect of reconnection is to peel off outer layers of the flux rope, thus decreasing MC thickness and reducing its magnetic flux, and consequently, affecting also the axial electric current. However, the presented order of magnitude considerations show that the effect of magnetic reconnection is at least one order of magnitude too weak to explain the noted discrepancies. From this, it can be concluded that the only viable effect that can provide an explanation for the observed MC evolutionary characteristics is the ``pancaking effect'' \citep[see, e.g.,][]{cargill94,cargill96,cargill99,mulligan01grl,mulligan01jgr,riley04,owens06}, i.e., the effect that leads to a deformation of a flux-rope that initially has a circular cross section, expanding much more in the direction perpendicular to the plane of the flux-rope axis than in radial direction \citep[for a discussion see][and references therein]{mostl09sph}. In cases where the ``pancaking effect'' is very pronounced, the standard methods of the magnetic field reconstruction are not appropriate, leading to internal inconsistencies of the results, as demonstrated, e.g., numerically by \cite{riley04} and observationally by \cite{mulligan01jgr}.
  In the latter paper the effect of MC ``flattening'' was inferred by studying data from two spacecraft (PVO and ISEE 3), separated longitudinally by 0.21 AU  and radially by 0.02 AU. The analysis showed that fitting a non-cylindrical flux rope model to the observational data from both spacecraft simultaneously, results in a stretched rope having almost twice as much magnetic flux than estimated by the independent cylindrically symmetric fit at PVO and five times larger than the flux calculated using the ISEE 3 data. Thus, a deviation from the cylindrically symmetric approximation is the most probable explanation for the apparent flux decrease found in the events under study.

\acknowledgments

This work has been supported in part by Croatian Science Foundation under the project no. 7549 "Millimeter and submillimeter observations of the solar chromosphere with ALMA" (MSOC). M.D. acknowledges funding from the EU H2020 MSCA grant agreement No 745782 (project ForbMod). C.J.F. and A.F.G. acknowledge that work at UNH was supported by NASA Wind grant NNX16Ao04G and NASA STEREO Quadrature grants.
 TA, CM and AV thank the Austrian Science Fund (FWF): [P31265-N27], [P26174-N27] and [P27292-N20]. We are thankful to the referee for thoughtful comments, which lead to a significant improvement of the paper.

\appendix

\section{Gold-Hoyle Configuration}

In the force-free uniform-twist  Gold-Hoyle configuration \citep[][hereafter GHC]{gold60} the axial and poloidal magnetic field components are defined as:
\begin{equation}
     B_{\parallel}(\tilde r) = \frac{B_c}{1+X^2{\tilde r}^2}\,,
\end{equation}
\begin{equation}
     B_{\phi}(\tilde r) = \frac{B_c X \tilde r}{1+X^2{\tilde r}^2}\,,
\end{equation}
respectively, where $\tilde r$ is the radial coordinate normalized to the flux-rope minor radius $r$, $B_c$ is the magnetic field at the rope axis, and $X$ is field-line twist per unit length:
\begin{equation}\label{E:X}
    X = \phi r/l  = 2\pi N\zeta\,,
\end{equation}
where $\phi$ is end-to-end twist ("total twist"), $l$ is the end-to-end length of the rope, and $N=2\pi/\phi$ is the number of turns a field line makes from one end of the flux rope to another and we abbreviated $\zeta=r/l$. In the GHC $\phi$ and $N$ do not depend on $\tilde r$, meaning by definition that it represents a ``uniform twist'' configuration. Note that both $\phi$ and $N$ are constant, due to the photospheric line-tying condition. The parameter $X$ can be also expressed as:
\begin{equation}
    X = \left(\frac{B_{\phi}}{B_{\parallel}}\right)_{\tilde r=1} = \tan\vartheta_{\tilde r=1}\,,
\end{equation}
where $\vartheta$ is the pitch angle of the field line (note that in the uniform twist case, $\tan\vartheta\propto \tilde r$). Thus, $X$ represents the tangent of the field-line pitch angle at the flux-rope surface, or equivalently, the ratio $B_{\phi}/B_{\parallel}$ at the flux-rope surface.

Integrating $B_z(\tilde r)$ over the flux-rope cross section, one finds the total longitudinal flux of the GHC:
\begin{equation}\label{E:Phi}
	 \Phi_{\parallel} = B_c\,\frac{\,r^2\pi}{X^2}\, \ln(1+X^2)=const.\,,
\end{equation}
i.e.,
\begin{equation}\label{E:Bc(Phi)}
	 B_c = \frac{C_1}{l^2\ln(1+\phi^2\zeta^2)}\,,
\end{equation}
where $C_1=\Phi_{\parallel}\phi^2/\pi=const.$, and $\zeta=r/l$.

On the other hand, employing the relation $I = 2r\pi (B_{\phi})_{\tilde r=1}$, one finds the total axial current of the GHC:
\begin{equation}
	I = B_c\,\frac{2\pi}{\mu}\,\frac{rX}{1+X^2} = B_c\,l\,\frac{C_2\,\zeta^2}{1+\phi^2\zeta^2}\,,
\end{equation}
where $C_2 = (2\pi\phi/\mu) = const.$ From this one finds:
\begin{equation}\label{E:Bc(I)}
	B_c = \frac{1+\phi^2\zeta^2}{C_2\,\zeta^2}\, \frac{\kappa_I}{l^2}\,,
\end{equation}
where we have taken into account $I\propto 1/l$ (see Section~\ref{S:concept}), i.e., $I(R)=\kappa_I/l(R)$. The constant $\kappa_I$ can be expressed as $\kappa_I=l_1I_1$, where $l_1$ and $I_1$ are the flux-rope length and current for the MC at 1\,AU.

Equating Eq.~(\ref{E:Bc(Phi)}) and Eq.~(\ref{E:Bc(I)}), one finds:
\begin{equation}
	 \frac{\zeta^2\,\ln(1+\phi^2\zeta^2)}{1+\phi^2\zeta^2} = \frac{C_1C_2}{\kappa_I} = const.\,,
\end{equation}
which can be satisfied only if $\zeta=const.$, since $\phi=const.$
 Bearing this in mind, Eq.~(\ref{E:X}) and Eq.~(\ref{E:Bc(Phi)}) imply $\vartheta=const.$ and $B_c\propto 1/l^2$, respectively, whereas $r/l\equiv\zeta=const.$ implies $r\propto l$, as well as $\Phi_{\parallel}=const.$. To summarize, the flux-rope thickness, $d=2r$, the axial current, $I$, the central magnetic field, $B_c$, and the field-line pitch angle, can be expressed as:
\begin{equation}\label{E:f(l)}
	 d=\kappa_d\,l, ~~~I=\kappa_I\,l^{-1}, ~~~B_c=\kappa_B\,l^{-2},~~{\rm and}~~\vartheta=const.\,,
\end{equation}
with $\kappa_d=d_1/l_1$, $\kappa_I=I_1l_1$, and $\kappa_B=B_{c1}l_1^2$, where subscript ``1'' denotes values at 1\,AU. The last relationship in Eq.~(\ref{E:f(l)}) implies also $B_c\propto1/r^2$. Representing the overall shape of MC by a semi-toroidal flux rope, or some other shape satisfying $l\propto R$ (see Appendix B), the relationships defined in Eq.(\ref{E:f(l)}) can be rewritten as:
\begin{equation}\label{E:f(R)}
	 d=\kappa_d\,R, ~~~I=\kappa_I\,R^{-1}, ~~~B_c=\kappa_B\,R^{-2},~~{\rm and}~~\vartheta=const.\,.
\end{equation}
Such a behavior is usually qualified as ``self-similar expansion''.

\section{Flux-Rope Length\,--\,Distance Relationship}

\begin{figure}
\epsscale{0.70}
\plotone{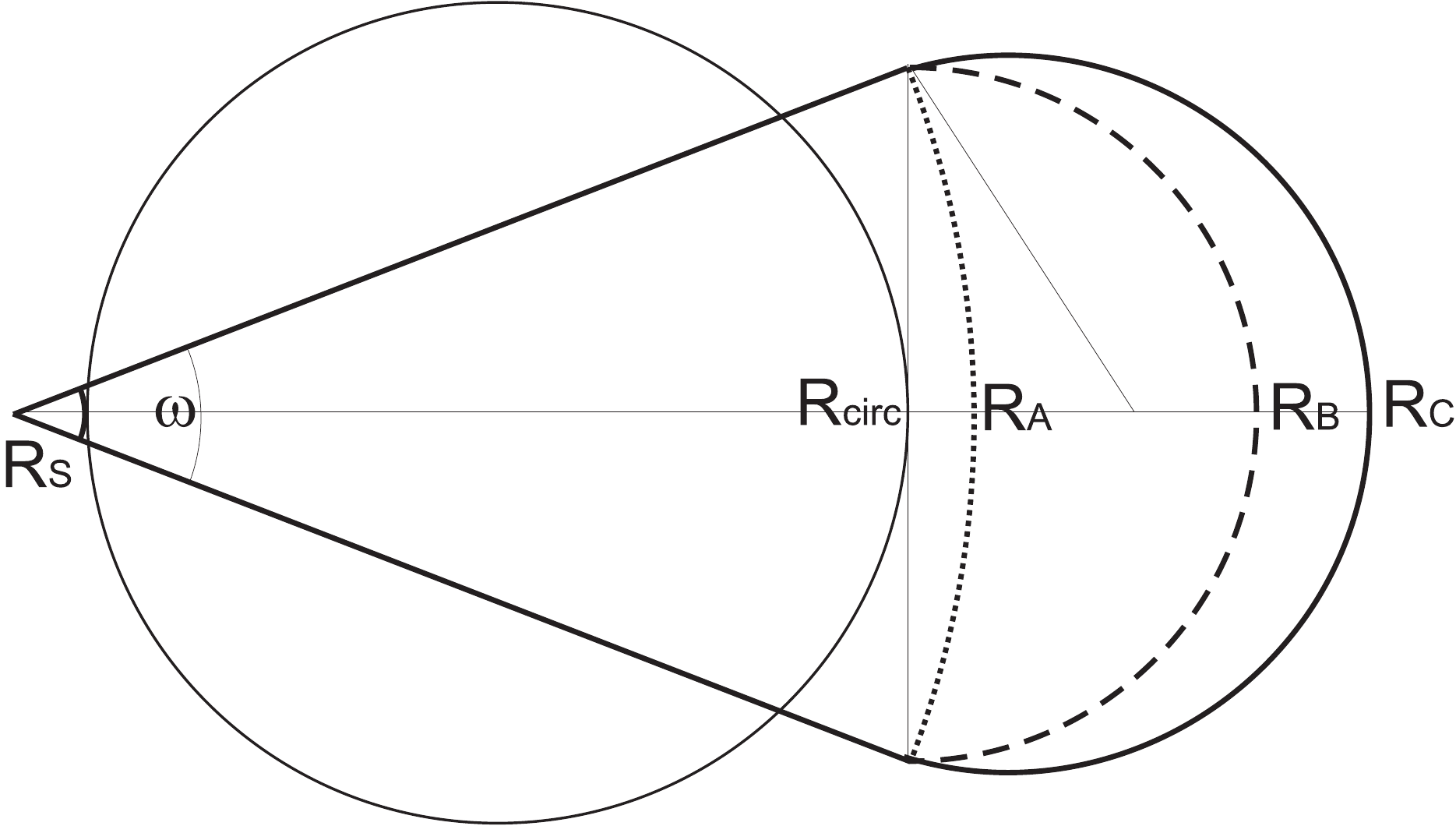}
  \caption
  {Sketch of the four considered shapes of the flux-rope axis (circular, cone-A, cone-B, and cone-C), whose summits reached the heliocentric distances $R_{circ}$, $R_A$, $R_B$, and $R_C$, respectively. Solar radius is denoted as $R_S$.
  \label{Af1}
  }
\end{figure}

In the following, the relationship between the flux-rope axis length, $l$, and the heliocentric distance, $R$, is considered for different geometries: 1) circular; 2) cone-A; 3) cone-B; 4) cone-C, which are shown in Figure~\ref{Af1}.
 It should be mentioned that the shapes cone-A and cone-B are not really appropriate to represent the flux-rope axis since they have ``knees'' at the points where the two radial lines connect to the circular frontal arc (in cone-A it is an arc concentric with the solar surface, and in cone-B it is a semi-circle).

For the mentioned flux-rope axis shapes the $l(R)$ relationships read:
\begin{equation}
    l = (R-R_S)\,\pi\,,
\end{equation}
\begin{equation}
    l = R\,(2+\omega) - R_S\,(2-\omega)\,,
\end{equation}
\begin{equation}
    l = R\left[\frac{2+\pi\sin\frac{\omega}{2}}{\cos\frac{\omega}{2}+\sin\frac{\omega}{2}}\right] - R_S\,(2-\omega)\,,
\end{equation}
\begin{equation}
    l = R\left[\frac{\left[2+(\pi+\omega)\tan\frac{\omega}{2}\right]\cos\frac{\omega}{2}}{1+\sin\frac{\omega}{2}}\right] - R_S\,(2-\omega)\,,
\end{equation}
respectively, where $R_S$ is the solar radius, and $\omega$ is the angle between the flux-rope axis legs. Note that Equations~(A1)\,--\,(A4) can be written in the form $l=aR-b$, where $a$ and $b$ are constants ($a_1=\pi$, $a_2=2+\omega$, $a_3=[2+\pi\sin(\omega/2)]/[\cos(\omega/2)+\sin(\omega/2)]$, $a_4=\cos(\omega/2)[2+(\pi+\omega)\tan(\omega/2)]/[1+\sin(\omega/2)]$, and $b_1=R_S\pi$, $b_2=b_3=b_4=R_S(2-\omega)$, respectively.

\begin{figure}
\epsscale{0.70}
\plotone{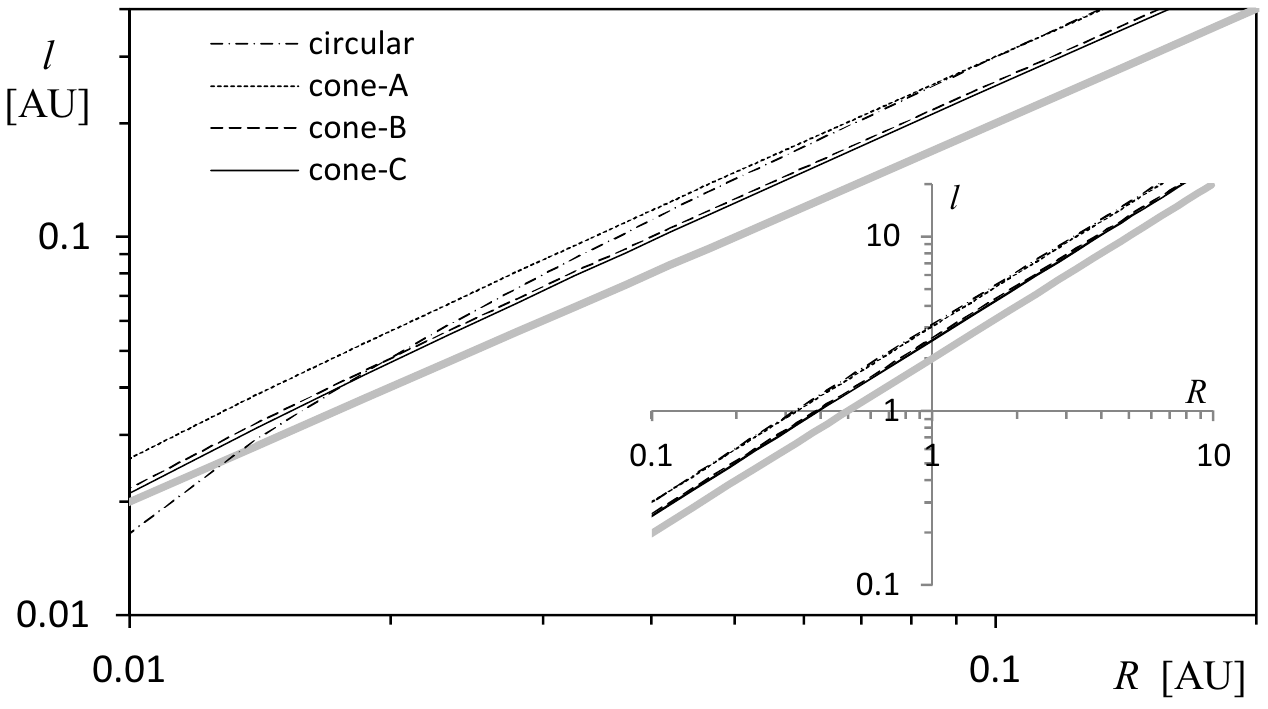}
  \caption
  {Flux-rope axis length versus heliospheric distance for different axis shapes in the range $R=0.01$\,--\,0.2 AU, corresponding to $R\approx 2$\,--\,40 solar radii. The dependence $l=2R$ (thick-gray line) is drawn to depict the $l\propto R$ slope. The extended range, $R=0.1$\,--\,10 AU (i.e., $R>20R_S$), where the departure from $l\propto R$ is below 0.1\,\%, is shown in the inset.
  \label{Af2}
  }
\end{figure}

In Figure~\ref{Af2} the dependencies $l(R)$ defined by Equations~(A1)\,--\,(A4) are displayed. The main graph represents the range 0.01\,AU\,$\leq R\leq$\,0.2\,AU (i.e., $R\sim$\,2\,--\,40\,$R_S$), where the deviations from the $l\propto R$ are significant. At heliocentric distances beyond 20\,$R_S$ the deviation from $l\propto R$ is negligible (see the graph in the inset of Figure~\ref{Af2}). For example, if the functions defined by Equations~(A1)\,--\,(A4) are fitted by the power-law ($l\propto R^k$) over a distance range 0.6\,--\,2.5 that is covered by measurements employed in this paper, the power-law slopes are $k=$ 1.0042, 1.0013, 1.0015, and 0.0016, respectively, i.e., the deviation from $k=1$ is on the order of 0.1\%. The difference becomes $\sim$\,1\,\% if the considered distance range is extended down to 10\,$R_S$.

\clearpage

\begin{table}[htbp]
  \caption{The event list; for details see the main text. In Column 6 the event labels from the sample used by \citet{leitner07} are displayed, together with labels (written in brackets) from \citet{farrugia05}.}
    \begin{tabular}{cccccccccc}
\hline
event	&		&		&	time range	&	distance	&		\\
label	&	year	&	data source	&	(DOY)	&	range (AU)	&	label*	\\
\hline
1	&	1974	&	IMP8, P11	&	285-299	&	1.00-4.80	&	1	\\
2	&	1975	&	H1, IMP8	&	321-321	&	0.87-1.00	&	2 (2)	\\
3	&	1977	&	H2, IMP8, V1	&	328-333	&	0.62-1.58	&	4	\\
4	&	1978	&	H2, OMNI, V1	&	004-008	&	0.94-1.98	&	5 (1)	\\
5	&	1978	&	H1, V2	&	060-069	&	0.87-2.49	&	6	\\
6	&	2009	&	MES, Wind	&	069-071	&	0.51-1.00	&	--	\\
7	&	2009	&	VEX, STA	&	191-193	&	0.73-0.96	&	--	\\
8	&	2010	&	MES, STB	&	309-313	&	0.47-1.08	&	--	\\
9	&	2011	&	VEX, STB	&	359-361	&	0.73-1.08	&	--	\\
10	&	2013	&	VEX, STA	&	008-010	&	0.72-0.96	&	--	\\
11	&	2013	&	MES, Wind	&	192-195	&	0.57-1.00	&	--	\\

\hline
    \end{tabular}
  \label{tab1}
\end{table}

\clearpage

\begin{table}[htbp]
  \caption{Basic data on the analyzed MC flux ropes obtained applying fitting to the Gold-Hoyle configuration. }
    \begin{tabular}{cccccccccccccc}
\hline
event	&	$R$ 	&	$v$ 	&	$\Delta t$	&	$B_{max}$ 	&	 $\phi$ 	&	 $\theta$	&	$p$	 & $d$  &	$B_c$ 	&	$H$	&	 $rms$	&	$E_{rms}$ 	\\
      	&	[AU]	&	[km\,s$^{-1}$]	&	 [h]	&	[nT]	&	[deg] 	&	[deg] 	&		&  [AU]  &	[nT]	&		&		&		\\
\hline
1a	&	1	&	449	&	31.9	&	18.7	&	97.2	&	-49.1	&	0.20	&	0.351	&	24.8	&	-1	&	5.81	&	0.31	\\
1b	&	4.8	&	418	&	45.0	&	5.7	&	95.4	&	-44.0	&	0.04	&	0.452	&	5.6	&	-1	&	3.03	&	0.53	\\
2a	&	0.87	&	330	&	9.0	&	14.4	&	84.1	&	-11.1	&	-0.12	&	0.071	&	15.0	&	-1	&	2.99	&	0.21	\\
2b	&	1	&	361	&	17.4	&	15.4	&	83.9	&	-28.0	&	-0.26	&	0.156	&	24.7	&	-1	&	2.69	&	0.18	\\
3a	&	0.62	&	289	&	9.3	&	36.8	&	121.9	&	12.4	&	-0.04	&	0.056	&	26.7	&	-1	&	13.28	&	0.36	\\
3b	&	1	&	462	&	18.6	&	18.1	&	40.8	&	-2.9	&	-0.01	&	0.135	&	16.5	&	-1	&	6.61	&	0.37	\\
3c	&	1.58	&	376	&	14.0	&	11.6	&	85.7	&	18.6	&	0.15	&	0.128	&	9.2	&	-1	&	3.88	&	0.33	\\
4a	&	0.94	&	526	&	28.5	&	20.9	&	257.7	&	44.3	&	0.20	&	0.363	&	22.8	&	1	&	5.61	&	0.27	\\
4b	&	1	&	581	&	36.0	&	19.9	&	230.1	&	5.1	&	0.02	&	0.387	&	22.5	&	1	&	7.67	&	0.39	\\
4c	&	1.98	&	579	&	52.0	&	10.2	&	260.5	&	45.0	&	0.25	&	0.743	&	7.7	&	1	&	2.80	&	0.28	\\
5a	&	0.87	&	443	&	33.2	&	27.9	&	79.9	&	54.9	&	-0.11	&	0.354	&	27.4	&	-1	&	7.82	&	0.28	\\
5b	&	2.49	&	469	&	72.0	&	6.0	&	73.0	&	9.2	&	0.00	&	0.779	&	5.9	&	-1	&	1.40	&	0.23	\\
6a	&	0.51	&	310	&	4.8	&	20.8	&	62.9	&	43.7	&	-0.01	&	0.034	&	18.0	&	1	&	6.22	&	0.30	\\
6b	&	1	&	353	&	17.5	&	15.6	&	91.7	&	79.0	&	0.01	&	0.149	&	10.2	&	1	&	3.71	&	0.24	\\
7a	&	0.73	&	290	&	24.8	&	15.8	&	230.1	&	20.4	&	-0.02	&	0.138	&	15.2	&	-1	&	4.47	&	0.28	\\
7b	&	0.96	&	315	&	13.7	&	8.5	&	242.7	&	17.3	&	0.11	&	0.094	&	8.3	&	-1	&	2.03	&	0.24	\\
8a	&	0.47	&	400	&	21.2	&	54.6	&	88.5	&	-60.1	&	-0.18	&	0.208	&	49.8	&	1	&	13.27	&	0.24	\\
8b	&	1.08	&	399	&	30.3	&	17.7	&	82.8	&	-32.2	&	-0.07	&	0.290	&	20.3	&	1	&	4.46	&	0.25	\\
9a	&	0.73	&	550	&	10.6	&	19.7	&	119.1	&	-6.8	&	-0.03	&	0.123	&	18.4	&	1	&	5.73	&	0.29	\\
9b	&	1.08	&	352	&	14.7	&	15.0	&	117.6	&	-13.4	&	-0.05	&	0.111	&	12.4	&	1	&	2.36	&	0.16	\\
10a	&	0.72	&	600	&	30.2	&	30.9	&	259.4	&	74.7	&	0.41	&	0.475	&	32.7	&	1	&	6.98	&	0.23	\\
10b	&	0.96	&	451	&	30.5	&	18.4	&	259.6	&	80.8	&	0.33	&	0.350	&	18.7	&	1	&	3.84	&	0.21	\\
11a	&	0.45	&	450	&	20.4	&	49.0	&	282.7	&	-7.9	&	0.01	&	0.215	&	44.7	&	-1	&	17.96	&	0.37	\\
11b	&	1	&	407	&	42.0	&	16.4	&	265.9	&	-21.7	&	0.05	&	0.411	&	14.5	&	-1	&	4.75	&	0.29	\\

\hline
    \end{tabular}
  \label{tab1b}
\end{table}

\clearpage

\begin{table}[htbp]
\scriptsize
  \centering
  \caption{Radial dependence of MC diameter, $d$, and central magnetic field, $B_c$,
  of the analyzed MCs, presented in a linear and power-law form. The extreme values obtained as illustrated in Figure~\ref{f1} are written as superscripts and subscripts, respectively. The average values and standard deviations are displayed in rows denoted as ``aver'' and ``stdev'', respectively. Note that E2 is excluded from calculating the mean values and standard deviations. }
    \begin{tabular}{cccccccccccc}
    \hline
    \multicolumn{ 1}{c}{} & \multicolumn{ 2}{c}{$d=a_dR+b_d$} &       & \multicolumn{ 2}{c}{$d=d_1\,R^{\alpha_d}$} &       & \multicolumn{ 2}{c}{$B_c=a_BR+b_B$} &       & \multicolumn{ 2}{c}{$B_c=B_{c1}\,R^{\alpha_B}$} \\
    \multicolumn{ 1}{c}{Event} & $a_d$ & $b_d$ &       & $d_1$ & $\alpha_d$ &       & $a_B$ & $b_B$ &       & $B_{c1}$ & $\alpha_B$ \\
    \hline
    1     & $0.03^{0.13}_{0.01}$ & $0.32^{0.16}_{0.35}$  &       & $0.35^{0.29}_{0.36}$  & $0.16^{0.64}_{0.05}$  &
                                          & $-5.03^{-4.49}_{-5.25}$   & $29.8^{27.2}_{30.1}$ &       & $24.8^{22.7}_{24.8}$  & $-0.94^{-0.89}_{-1.04}$ \\
    (2)   & $0.65^{0.69}_{0.14}$ & $-0.49^{-0.53}_{0.00}$ &      & $0.16^{0.16}_{0.14}$  & $5.60^{5.99}_{0.98}$  &
                                             & $74.7^{74.1}_{59.1}$  & $-50.0^{-49.4}_{-35.3}$ &       & $24.7^{24.6}_{23.7}$  & $3.58^{3.56}_{2.81}$  \\
    3     & $0.07^{0.16}_{0.06}$ & $0.03^{0.07}_{0.02}$  &       & $0.10^{0.08}_{0.08}$  & $0.90^{1.96}_{0.80}$  &
                                          & $-17.9^{-17.1}_{-18.9}$   & $36.5^{37.3}_{39.1}$ &       & $15.8^{16.5}_{15.7}$   & $-1.15^{-1.01}_{-1.16}$ \\
    4     & $0.36^{0.48}_{0.22}$ & $0.02^{-0.18}_{0.20}$ &       & $0.39^{0.30}_{0.42}$  & $0.96^{1.39}_{0.59}$  &
                                            & $-14.8^{-12.2}_{-14.8}$   & $37.0^{32.2}_{36.9}$ &       & $21.6^{19.1}_{21.0}$  & $-1.51^{-1.28}_{-1.48}$ \\
    5     & $0.26^{0.46}_{0.25}$ & $0.13^{-0.04}_{0.15}$ &       & $0.39^{0.41}_{0.40}$  & $0.75^{1.08}_{0.70}$  &
                                          & $-13.3^{-13.2}_{-13.4}$  & $38.9^{38.9}_{39.1}$  &       & $22.4^{22.4}_{22.3}$  & $-1.46^{-1.45}_{-1.49}$ \\
    6     & $0.23^{0.27}_{0.23}$ & $0.09^{0.11}_{0.08}$  &       & $0.15^{0.16}_{0.15}$  & $2.19^{2.47}_{2.05}$  &
                                          & $-15.9^{-15.7}_{-16.5}$  & $26.1^{26.3}_{26.7}$  &       & $10.2^{10.5}_{10.2}$  & $-0.84^{-0.82}_{-0.86}$ \\
    7   & $-0.19^{-0.13}_{-0.31}$ & $0.28^{0.22}_{0.38}$ &     & $0.09^{0.10}_{0.08}$  & $-1.42^{-0.96}_{-2.23}$ &
                                          & $-29.8^{-27.4}_{-30.0}$  & $36.9^{34.7}_{37.1}$  &       & $7.6^{7.7}_{7.6}$  & $-2.19^{-2.05}_{-2.21}$ \\
    8     & $0.13^{0.16}_{0.10}$ & $0.14^{0.12}_{0.17}$  &       & $0.28^{0.28}_{0.28}$  & $0.40^{0.48}_{0.29}$  &
                                          & $-48.4^{-46.4}_{-50.4}$  & $72.6^{70.4}_{74.6}$ &       & $22.0^{22.0}_{22.0}$  & $-1.08^{-1.05}_{-1.11}$ \\
    9   & $-0.03^{-0.02}_{-0.07}$ & $0.15^{0.14}_{0.19}$ &     & $0.11^{0.12}_{0.11}$  & $-0.26^{-0.16}_{-0.53}$ &
                                          & $-17.0^{-15.6}_{-17.8}$  & $30.8^{29.1}_{31.3}$  &       & $13.4^{13.2}_{13.1}$  & $-1.00^{-0.94}_{-1.07}$ \\
   10   & $-0.52^{-0.41}_{-0.68}$ & $0.85^{0.75}_{1.00}$ &     & $0.33^{0.35}_{0.32}$  & $-1.07^{-0.85}_{-1.36}$ &
                                          & $-58.3^{-56.3}_{-57.7}$  & $74.7^{72.8}_{75.9}$  &       & $17.3^{17.3}_{17.2}$  & $-1.94^{-1.89}_{-1.99}$ \\
   11     & $0.45^{0.57}_{0.42}$ & $-0.04^{-0.12}_{-0.01}$ &     & $0.41^{0.44}_{0.41}$  & $1.15^{1.42}_{1.04}$  &
                                          & $-70.2^{-70.7}_{-73.2}$  & $84.7^{85.3}_{87.6}$  &       & $14.5^{14.6}_{14.5}$  & $-2.01^{-2.00}_{-2.06}$ \\
    \hline
    aver  & $0.08^{0.17}_{0.02}$  & $0.20^{0.12}_{0.25}$  &       & $0.26^{0.25}_{0.26}$  & $0.38^{0.74}_{0.14}$  &
                                           & $-29.1^{-27.9}_{-29.8}$  & $46.8^{45.4}_{47.8}$  &       & $17.0^{16.6}_{16.8}$ & $-1.41^{-1.34}_{-1.45}$ \\
    stdev & $0.28^{0.30}_{0.32}$  & $0.25^{0.25}_{0.29}$  &       & $0.13^{0.13}_{0.14}$  & $1.08^{1.14}_{1.24}$  &
                                           & $22.1^{22.1}_{22.7}$  & $21.6^{21.9}_{22.4}$  &       & $5.7^{5.1}_{5.7}$ & $0.49^{0.48}_{0.48}$ \\
    \hline
    \end{tabular}
  \label{tab2}
\end{table}

    \clearpage

\begin{table}[htbp]
\scriptsize
  \centering
  \caption{Radial dependence of
   the inferred axial electric current, $I_{\parallel}$, and magnetic flux, $\Phi_{\parallel}$, of the analyzed MCs, presented in a linear and power-law form. The extreme values obtained as illustrated in Figure~\ref{f1} are written as superscripts and subscripts, respectively. The average values and standard deviations are displayed in rows denoted as ``aver'' and ``stdev'', respectively. Note that E2 is excluded from calculating the mean values and standard deviations. }
    \begin{tabular}{cccccccccccc}

    \hline
    \multicolumn{ 1}{c}{} & \multicolumn{ 2}{c}{$I_{\parallel}=a_IR+b_I$} &       & \multicolumn{ 2}{c}{$I_{\parallel}=I_1\,R^{\alpha_I}$} &       & \multicolumn{ 2}{c}{$\Phi_{\parallel}=a_{\Phi}R+b_{\Phi}$} &       & \multicolumn{ 2}{c}{$\Phi_{\parallel}=\Phi_1\,R^{\alpha_{\Phi}}$} \\
    \multicolumn{ 1}{c}{Event} & $a_I$ & $b_I$ &       & $I_1$ & $\alpha_I$ &       & $a_{\Phi}$ & $b_{\Phi}$ &       & $\Phi_1$ & $\alpha_{\Phi}$ \\
    \hline
    1     & $-0.25^{-0.23}_{-0.25}$  & $1.40^{1.28}_{1.39}$ &       & $1.15^{1.05}_{1.14}$  & $-1.15^{-1.07}_{-1.20}$  &
                                             & $-0.39^{-0.31}_{-0.39}$  & $1.15^{1.07}_{1.20}$ &       & $1.77^{1.51}_{1.76}$  & $-1.18^{-0.96}_{-1.18}$ \\
    (2) & $2.46^{2.48}_{1.32}$  & $-1.94^{-1.96}_{-0.84}$ &       & $0.52^{0.52}_{0.48}$  & $6.84^{7.02}_{3.20}$  &
                                            & $2.05^{2.09}_{0.96}$  & $-1.69^{-1.73}_{-0.65}$ &       & $0.36^{0.36}_{0.30}$  & $9.67^{10.1}_{3.77}$ \\
    3     & $-0.08^{-0.17}_{0.14}$  & $0.38^{0.01}_{0.36}$ &      & $0.28^{0.18}_{0.19}$  & $-0.26^{0.96}_{-0.73}$  &
                                             & $0.08^{0.21}_{0.10}$  & $0.12^{-0.09}_{0.05}$ &       & $0.19^{0.10}_{0.15}$  & $0.66^{1.98}_{0.66}$ \\
    4     & $-0.52^{-0.29}_{-0.63}$  & $1.76^{1.31}_{1.93}$ &       & $1.23^{1.01}_{1.26}$  & $-0.74^{-0.47}_{-0.92}$  &
                                             & $0.15^{0.69}_{-0.58}$  & $2.09^{1.03}_{3.02}$ &       & $2.24^{1.73}_{2.42}$  & $0.09^{0.48}_{-0.38}$ \\
    5     & $-0.50^{-0.50}_{-0.52}$  & $1.86^{1.86}_{1.87}$ &       & $1.27^{1.27}_{1.27}$  & $-0.80^{-0.80}_{-0.84}$  &
                                             & $-0.16^{0.01}_{-0.19}$   & $2.50^{2.35}_{2.54}$  &    & $2.32^{2.36}_{2.33}$  & $-0.11^{0.00}_{-0.13}$  \\
    6     & $0.35^{0.48}_{0.32}$  & $-0.06^{-0.17}_{-0.04}$ &       & $0.28^{0.31}_{0.28}$  & $1.36^{2.08}_{1.21}$  &
                                            & $0.49^{0.56}_{0.49}$  & $-0.23^{-0.26}_{-0.22}$  &       & $0.27^{0.30}_{0.27}$   & $3.41^{3.56}_{3.36}$ \\
    7     & $-0.89^{-0.82}_{-0.92}$  & $1.00^{0.94}_{1.02}$ &       & $0.13^{0.14}_{0.11}$  & $-3.24^{-2.93}_{-3.54}$  &
                                           & $-0.65^{-0.58}_{-0.73}$   & $0.73^{0.66}_{0.79}$  &       & $0.08^{0.09}_{0.07}$   & $-3.43^{-2.95}_{-4.04}$  \\
    8     & $-1.51^{-1.44}_{-1.56}$  & $2.57^{2.50}_{2.62}$ &       & $1.00^{1.00}_{0.99}$  & $-0.82^{-0.80}_{-0.85}$  &
                                           & $-1.36^{-1.15}_{-1.53}$   & $2.81^{2.59}_{2.98}$  &       & $1.40^{1.40}_{1.39}$   & $-0.58^{-0.50}_{-0.64}$  \\
    9     & $-0.40^{-0.43}_{-0.47}$  & $0.69^{0.71}_{0.75}$ &       & $0.28^{0.27}_{0.28}$  & $-1.12^{-1.19}_{-1.28}$  &
                                          & $-0.26^{-0.28}_{-0.39}$   & $0.46^{0.48}_{0.58}$  &       & $0.19^{0.19}_{0.18}$   & $-1.08^{-1.17}_{-1.55}$  \\
   10     & $-3.79^{-3.75}_{-3.83}$  & $4.79^{4.75}_{4.82}$ &       & $1.06^{1.06}_{1.05}$  & $-2.03^{-2.01}_{-2.05}$  &
                                           & $-8.78^{-8.37}_{-9.34}$   & $10.6^{10.3}_{11.1}$  &       & $1.99^{2.02}_{1.97}$   & $-2.34^{-2.24}_{-2.46}$  \\
   11     & $-1.46^{-1.11}_{-1.61}$  & $2.44^{2.12}_{2.58}$ &       & $0.97^{1.01}_{0.97}$  & $-0.89^{-0.69}_{-0.96}$  &
                                             & $0.58^{1.60}_{028}$   & $1.45^{0.59}_{1.76}$  &       & $2.03^{2.18}_{2.03}$   & $0.23^{0.67}_{0.11}$  \\
    \hline
    aver  & $-0.95^{-0.79}_{-0.96}$  & $1.68^{1.53}_{1.73}$ &        & $0.77^{0.73}_{0.75}$  & $-0.97^{-0.69}_{-1.11}$  &
                                             & $-1.03^{-0.76}_{-1.23}$  & $2.17^{1.87}_{2.38}$ &       & $1.25^{1.19}_{1.26}$  & $-0.43^{-0.11}_{-0.63}$ \\
    stdev & $1.17^{1.18}_{1.17}$  & $1.39^{1.42}_{1.40}$ &        & $0.41^{0.44}_{0.48}$  & $1.17^{1.40}_{1.18}$  &
                                             & $2.78^{2.78}_{2.91}$  & $3.13^{3.10}_{3.28}$ &       & $0.95^{0.92}_{0.98}$  & $1.84^{1.93}_{1.96}$ \\
    \hline
    \end{tabular}
  \label{tab3}
\end{table}

\clearpage

\begin{table}[htbp]
  \caption{Comparison of the increase rates of the MC diameter and the decrease rate of the magnetic field. For details see the main text. Note that E2 is excluded from calculating the mean values and standard deviations.}
    \begin{tabular}{ccccccccrcccr}
\hline
    Event &  $2\alpha_d$  &   $\alpha_B$  & $2\alpha_d-\alpha_B$ &  $\Delta$\%   & $\frac{2\alpha_d}{\alpha_B}$ && $1+\alpha_d$  & $\frac{1+\alpha_d}{-\alpha_B}$   \\
\hline
     1	  &  0.32	& $-0.94$	& $-0.62$	&  $66$ & $-0.34$  && 1.16  &  1.23 \\
    (2)   & 11.2    & 3.58	    & 14.78	    & $413$ & $3.13$   && 6.60  &  1.84  \\
    3     & 1.80    & $-1.15$   & 0.65      & $-57$ & $-1.57$  && 1.90  &  1.65 \\
    4     & 1.92    & $-1.51$   & 0.41      & $-27$ & $-1.27$  && 1.96  &  1.30 \\
    5     & 1.50    & $-1.45$   & 0.05      & $-3$  & $-1.03$  && 1.75  &  1.21 \\
    6     & 4.38    & $-0.84$   & 3.54      & $-421$ & $-5.21$ && 3.19  &  3.80 \\
    7     & $-2.84$ & $-2.19$   & $-5.03$   & $230$ & $1.30$   && $-0.42$  &  $-0.19$ \\
    8     & 0.80    & $-1.08$   & $-0.28$   & $26$  & $-0.74$  && 1.40  &  1.30 \\
    9     & $-0.52$ & $-1.00$   & $-1.52$   & $152$ & $0.52$   && 0.74  &  0.74 \\
    10    & $-2.14$ & $-1.94$   & $-4.08$   & $210$ & $1.10$   && $-0.07$  &  $-0.04$ \\
    11    & 1.15    & $-2.01$   & $-0.86$   & $43$  & $-0.57$  && 1.58  &  0.78 \\
\hline
    aver  & 0.64    & $-1.41$   & $-0.77$   & $22$  & $-0.78$  && 1.32  & 1.18  \\
    stdev & 2.09    & 0.49     & 2.42       & 184   & 1.84     && 1.04  & 1.10  \\
\hline
    \end{tabular}
  \label{tab4}
\end{table}

\clearpage

\begin{table}[htbp]
  \centering
  \caption{Observed change of the axial magnetic flux, $\Delta\Phi_{\rm obs}$, compared with the estimated reconnected flux, $\Delta\Phi_{\rm recon}$. $R_a$ and $R_b$ are heliocentric distances of the first and the last spacecraft measurement, respectively, whereas $\overline V_{MC}$ represents the mean MC speed over this distance range.}
    \begin{tabular}{ccccccc}
\hline
    event & $R_a$    & $R_b$    & $\overline V_{MC}$ & $\Delta\Phi_{\rm obs}$ & $\Delta\Phi_{\rm recon}$ & $\frac{\Delta\Phi_{\rm recon}}{\Delta\Phi_{\rm obs}}$ \\
            & AU  & AU & km\,s$^{-1}$ & $10^{21}$\,Mx  & $10^{21}$\,Mx   &  \\
\hline
    1     & 1.00  & 4.80 & 434  & 1.49  & 0.10  & 0.07 \\
    5     & 0.87  & 2.49 & 456  & 0.26  & 0.09  & 0.35 \\
    7     & 0.73  & 0.96 & 303  & 0.15  & 0.06  & 0.41 \\
    8     & 0.47  & 1.08 & 400  & 0.83  & 0.17  & 0.20 \\
    9     & 0.73  & 1.08 & 451  & 0.09  & 0.06  & 0.60 \\
    10    & 0.72  & 0.96 & 526  & 2.11  & 0.04  & 0.02 \\
\hline
    \end{tabular}
  \label{tab5}
\end{table}


\clearpage

\clearpage

\clearpage

\clearpage

\clearpage

\clearpage

\clearpage


\clearpage


\clearpage



\clearpage


\clearpage

\bibliographystyle{plainnat}
\bibliography{cme_bib}

\end{document}